\documentclass[12pt,preprint]{elsarticle} 
\usepackage[utf8]{inputenc}
\usepackage[a4paper,margin=1in]{geometry}
\usepackage{amsmath,amssymb,bm}
\usepackage{graphicx}
\usepackage{booktabs}
\usepackage{longtable}
\usepackage{enumitem}
\usepackage{xcolor}
\usepackage{hyperref}
\usepackage{mathtools}
\usepackage{microtype}
\usepackage{tikz}
\usepackage{subcaption}

\hypersetup{
    colorlinks=true,
    linkcolor=blue!60!black,
    citecolor=blue!60!black,
    urlcolor=blue!60!black
}

\begin{document}
\begin{frontmatter} 
\title{A Low-Storage Implicit Dual-Time Finite-Volume Framework for Radio-Frequency Capacitively Coupled Plasma Fluid Simulations}

\author[a]{Yuze ZHU}
\author[a]{Hangkong WU}
\author[a]{Junzhe CAO}
\author[a]{Yufeng WEI}
\author[a,b,c]{Kun XU\corref{cor1}} 
\cortext[cor1]{Corresponding author. Email: makxu@ust.hk}

\affiliation[a]{
   organization= {Department of Mathematics, Hong Kong University of Science and Technology},
   city= {Clear Water Bay, Kowloon, Hong Kong},
   country= {China},
}

\affiliation[b]{
   organization= {Department of Mechanical and Aerospace Engineering, Hong Kong University of Science and Technology},
   city= {Clear Water Bay, Kowloon, Hong Kong},
   country= {China},
}

\affiliation[c]{
   organization= {Shenzhen Research Institute, Hong Kong University of Science and Technology},
   city= {Shenzhen},
   country= {China},
}

\begin{abstract}
Radio-frequency (RF) capacitively coupled plasmas (CCPs) are widely utilized in semiconductor manufacturing. Efficiently and accurately solving the underlying fluid governing equations to resolve the complex multi-physics fields is crucial for optimizing plasma reactor designs and process control. To overcome the severe numerical stiffness and prohibitive time-step constraints inherent in low-temperature plasma modeling, we present a robust, low-storage implicit dual-time finite-volume framework for RF CCP simulations, establishing a highly efficient and memory-friendly pathway for the predictive modeling of multi-dimensional low-temperature plasmas. In this approach, the physical time advancement is strictly decoupled from explicit stability limits through a backward-difference formula (BDF), while the resulting nonlinear system is efficiently solved using pseudo-time iterations. A localized block-implicit relaxation method is employed to handle the stiff transport and chemical source terms at the cell level, effectively circumventing the massive memory overhead typical of conventional fully implicit solvers. Concurrently, a semi-implicit treatment of Poisson's equation is integrated to accelerate the electrostatic coupling. The framework is first verified against a standard one-dimensional argon discharge benchmark, demonstrating that a highly accurate periodic state can be achieved with satisfactory computational efficiency through the optimal selection of the physical time step, pseudo-CFL number, and inner iteration step. To further demonstrate the multidimensional applicability of the proposed method, the solver is extended to genuine two-dimensional configurations. The numerical results show the multi-dimensional distortion of the electrostatic potential and localized electron heating zones induced by the transverse boundaries, highlighting the solver's capability to resolve complex, genuinely two-dimensional plasma-sheath interactions.
\end{abstract}


\end{frontmatter}

\section{Introduction}

Driven by the demands of advanced semiconductor manufacturing, radio-frequency (RF) capacitively coupled plasmas (CCPs) are extensively deployed in low-temperature surface processing, facilitating key processes such as semiconductor etching and thin-film deposition~\cite{lieberman1994principles,makabe2006plasma,donnelly2013plasma}. The extensive deployment stems from their straightforward hardware configuration, stable discharge conditions, and the critical ability to effectively tune ion energy at the substrate~\cite{boyle2004independent,kitajima2000functional,heil2008possibility}. To further optimize these reactors for precise plasma control, a comprehensive understanding of the internal discharge dynamics is strictly required. While experimental diagnostics provide irreplaceable insights, they are often expensive, time-consuming, and hampered by probe intrusiveness or limited optical access. With the continuous development of computational resources, numerical simulation has evolved into a fundamental tool in plasma research. By providing high-resolution details, it plays an increasingly important role in deepening physical understanding and accelerating design iterations, particularly in reactor design and process optimization~\cite{kushner2009hybrid,economou2000modeling,verboncoeur2005particle,dawson1983particle}.

Despite the importance of numerical simulation in improving the plasma source performance, the time-domain simulation of RF CCPs remains computationally demanding. The discharge dynamics involve strong nonlinear coupling among macroscopic plasma transport, reaction kinetics, and the electrostatic field~\cite{chabert2011physics,raizer2017radio}. To capture these multiscale phenomena, two modeling paradigms are widely employed: kinetic and fluid models. While kinetic approaches, such as particle-in-cell/Monte Carlo collision (PIC/MCC) methods, provide detailed descriptions of non-equilibrium particle behavior ~\cite{birdsall2018plasma,vahedi1995monte}, their high computational cost typically limits their application in multidimensional, high-pressure, or chemically complex reactor-scale simulations. Alternatively, fluid models which solve macroscopic continuity and energy equations coupled with Poisson's equation offer a more computationally tractable approach~\cite{stewart1994two,graves1987fluid,lymberopoulos1993fluid}. However, fluid simulations of RF CCPs suffer from severe numerical stiffness due to the highly disparate time scales of electron transport, dielectric relaxation, and chemical reactions. In conventional explicit or semi-implicit schemes, the physical time step is strictly dictated by numerical stability conditions rather than the actual temporal resolution needed to capture the RF dynamics. This forces the use of excessively small time steps, making the long transient evolution toward a periodic steady state computationally prohibitive~\cite{sommerer1992numerical}.

To overcome these stringent stability constraints, various numerical strategies have been developed over the past few decades. Early efforts predominantly focused on semi-implicit treatments of Poisson's equation to relax the restrictive time-step limits imposed by dielectric relaxation~\cite{boeuf1988two}. Subsequently, Hagelaar and Kroesen proposed an implicit linearization of the electron-energy source term, which significantly alleviated the stiffness induced by highly nonlinear collision processes~\cite{hagelaar2000speeding}. To further improve global stability and manage the strong inter-equation coupling, more sophisticated approaches were introduced. These include implicit Newton--Krylov algorithms, often combined with massively parallel finite-volume implementations~\cite{hammond2002numerical,hammond2003numerical,lin2012development,chen2020development}. Furthermore, specialized algorithmic accelerations tailored specifically for RF CCP models have been actively explored, including multi-time-step explicit integration and periodic-convergence extrapolation techniques~\cite{li2025fast,gogolides1992direct}.

Although these advancements have substantially improved the efficiency of plasma fluid simulations, fundamental trade-offs persist among numerical stability, single-step computational cost, implementation complexity, and algorithmic generality. Conventional explicit and semi-explicit schemes offer low cost per iteration but remain vulnerable to the stiffness induced by electron transport and fast chemical reactions. Conversely,  implicit Newton--Krylov methods robustly address the nonlinear coupling between drift-diffusion processes and electrostatic fields, permitting large physical time steps. However, these methods require solving global nonlinear systems, computing intricate Jacobian (or Jacobian-vector) products, and designing highly specialized preconditioners. This inevitably leads to massive memory footprints and complicates the solver implementation. While massive parallelization reduces the overall wall-clock time, it does not fundamentally alleviate the underlying algorithmic stiffness. Therefore, there remains a strong need to develop a unified time-domain fluid solver that simultaneously achieves the robust stability of implicit coupling, the low memory requirements of localized relaxation methods, and the flexibility to accommodate complex reaction kinetics.

Motivated by these persisting challenges, this work proposes a low-storage, implicit dual-time finite-volume framework accommodated for RF CCP fluid simulations. By employing a backward-difference formula (BDF) to construct the physical-time residual, pseudo-time iterations are introduced at each physical time step. This dual-time formulation fundamentally decouples the physical time advancement from numerical stability restrictions, transforming the stiff unsteady update into a sequence of quasi-steady problems. To update the charged-particle and electron-energy equations without incurring massive memory costs, the discretized system is advanced through a localized block-implicit relaxation strategy. By resolving the stiff inter-equation coupling at the cell level, this approach effectively bypasses the assembly and inversion of a prohibitive global transport Jacobian. Concurrently, the electrostatic potential is advanced via a semi-implicit method to robustly manage the space-charge coupling. Ultimately, this framework introduces a novel and highly efficient alternative to traditional implicit solvers in plasma simulations, simultaneously achieving low memory consumption and high temporal accuracy. 
The proposed framework is verified against a classic one-dimensional RF argon CCP benchmark, implemented within a two-dimensional computational solver. Comprehensive numerical experiments are conducted to examine the impact of pseudo-time convergence criteria and physical time-step sizes on the overall computational efficiency in reaching the periodic steady state.

The remainder of this paper is organized as follows. Section 2 introduces the macroscopic plasma-fluid equations and the nondimensionalization scaling, before detailing the numerical implementation. This includes the dual-time finite-volume discretization, the localized block-implicit relaxation, and the semi-implicit treatment of Poisson's equation. Section 3 presents comprehensive numerical results and performance verification for both quasi-one-dimensional and genuine two-dimensional argon discharges, highlighting the  temporal accuracy, computational efficiency, and capability of the proposed framework to capture complex multi-dimensional sheath dynamics. Finally, conclusions are drawn in Section 4.

\section{Mathematical and Numerical Formulation}

\subsection{Governing Equations and Nondimensionalization}

In this work, the radio-frequency capacitively coupled plasma is described by a continuum plasma-fluid model. The characteristic reactor dimension considered here is much smaller than the electromagnetic wavelength associated with the applied RF frequency, and inductive electromagnetic effects are neglected. The electric field is therefore modeled using the electrostatic approximation as follows:
\begin{equation}
\mathbf{E}=-\nabla \phi,
\end{equation}
where $\phi$ is the electrostatic potential which is determined self-consistently from Poisson's equation:
\begin{equation}
-\nabla\cdot\left(\epsilon_0\nabla\phi\right)
=
e\left(n_i-n_e\right),
\label{eq:poisson_dim}
\end{equation}
where $n_e$ and $n_i$ are the electron and positive ion number densities, $e$ is the elementary charge, and $\epsilon_0$ is the permittivity of free space.

To capture the essential discharge dynamics, the model generally considers electrons, positive ions, and excited metastable species. Under the collisional conditions considered here, the charged and neutral species fluxes can be represented by the drift-diffusion approximation. The continuity equations for electrons, positive ions, and metastable atoms (denoted by the subscript $*$) are given by
\begin{equation}
\frac{\partial n_e}{\partial t}
+
\nabla\cdot\boldsymbol{\Gamma}_e
=
S_e,
\label{eq:electron_continuity}
\end{equation}
\begin{equation}
\frac{\partial n_i}{\partial t}
+
\nabla\cdot\boldsymbol{\Gamma}_i
=
S_i,
\label{eq:ion_continuity}
\end{equation}
\begin{equation}
\frac{\partial n_*}{\partial t}
+
\nabla\cdot\boldsymbol{\Gamma}_*
=
S_*.
\label{eq:meta_continuity}
\end{equation}

The particle fluxes consist of a drift term driven by the macroscopic electric field and a diffusion term driven by concentration gradients. Since metastable atoms are uncharged, their transport is purely diffusive. The respective fluxes are modeled as
\begin{equation}
\boldsymbol{\Gamma}_e
=
-\mu_e n_e \mathbf{E}
+
D_e\nabla n_e, 
\label{eq:electron_flux}
\end{equation}
\begin{equation}
\boldsymbol{\Gamma}_i
=
-\mu_i n_i \mathbf{E}
-
D_i\nabla n_i, 
\label{eq:ion_flux}
\end{equation}
\begin{equation}
\boldsymbol{\Gamma}_*
=
-D_*\nabla n_*,
\label{eq:meta_flux}
\end{equation}
where $\mu$ and $D$ represent the mobilities and diffusion coefficients of the respective species. 

The net source terms ($S_e$, $S_i$, and $S_*$) on the right-hand side of the continuity equations account for gas-phase reactions. For charged particles, the generation mechanisms typically involve direct ionization from the ground state, stepwise ionization from the metastable state, and metastable pooling reactions. Consequently, the electron and ion source terms are essentially equivalent ($S_e = S_i$) and can be expanded as
\begin{equation}
S_e
=
k_{\rm ion} N_g n_e 
+
k_{\rm si} n_* n_e
+
k_{\rm mp} n_*^2,
\label{eq:source_terms}
\end{equation}
where $N_g$ is the neutral argon gas density, and $k_{\rm ion}$, $k_{\rm si}$, and $k_{\rm mp}$ are the rate coefficients for direct ionization, stepwise ionization, and metastable pooling, respectively.

The electron energy equation is written in conservative form to account for energy transport, Joule heating, and collisional energy losses:
\begin{equation}
\frac{\partial \varepsilon_e}{\partial t}
+
\nabla\cdot\boldsymbol{\Gamma}_{\varepsilon}
=
\mathbf{J}_e\cdot\mathbf{E}
-
\sum_{j} \Delta\mathcal{E}_{j} S_{j},
\label{eq:electron_energy}
\end{equation}
where $\varepsilon_e = \frac{3}{2} n_e e T_e$ represents the electron energy density, with $T_e$ denoting the electron temperature in electron-volts ($\mathrm{eV}$), and $\mathbf{J}_e=-e\boldsymbol{\Gamma}_e$ is the electron current density. The electron energy flux, denoted by $\boldsymbol{\Gamma}_{\varepsilon}$, is modeled under the drift-diffusion approximation as
\begin{equation}
\boldsymbol{\Gamma}_{\varepsilon}
=
-\frac{5}{3}\mu_e\varepsilon_e\mathbf{E}
-
\frac{5}{3}D_e\nabla\varepsilon_e. 
\label{eq:energy_flux}
\end{equation}

On the right-hand side of Eq.~\eqref{eq:electron_energy}, the first term, $\mathbf{J}_e\cdot\mathbf{E}$, represents the electron Joule heating. The summation term encompasses the total inelastic energy losses across various collision channels, such as direct ionization and ground-state excitation. For each $j$-th inelastic collision process, $\Delta\mathcal{E}_j$ dictates the specific threshold energy loss, and $S_j$ represents its volumetric reaction rate. 

Because the rate coefficients for these electron-impact collisions depend strongly on the electron energy, they are typically tabulated as functions of the mean electron energy. In the present macroscopic fluid framework, this mean electron energy $\bar{\varepsilon}_e$ is defined directly from the transported variables as
\begin{equation}
\bar{\varepsilon}_e
=
\frac{\varepsilon_e}{n_e}
=
\frac{3}{2} e T_e.
\end{equation}

To mitigate numerical stiffness arising from the vast disparity in the magnitudes of physical quantities, the state variables and governing equations are nondimensionalized.
Let $L_{\rm ref}$, $n_{\rm ref}$, $T_{e,\rm ref}$, and $\Phi_{\rm ref}$ denote the chosen reference length, number density, electron temperature, and electric potential, respectively. Based on these primary quantities, the reference electron thermal speed and the reference time are defined as

\begin{equation}
v_{\rm ref}
=
\left(
\frac{8 k_B T_{e,\rm ref}}{\pi m_e}
\right)^{1/2},
\label{eq:vref_tref}
\end{equation}

\begin{equation}
t_{\rm ref}
=
\frac{L_{\rm ref}}{v_{\rm ref}},
\label{eq:tref}
\end{equation}
where $k_B$ is the Boltzmann constant and $m_e$ is the electron mass. The reference electron energy density is scaled by
\begin{equation}
\varepsilon_{\rm ref}
=
\frac{3}{2}
n_{\rm ref} k_B T_{e,\rm ref}.
\label{eq:eref}
\end{equation}

The scaling factors for all other secondary variables, including transport coefficients, reaction rate coefficients, and vacuum permittivity are derived consistently from these primary reference values. By setting up the reference scales in this manner, the dimensionless governing equations retain the exact form of their dimensional counterparts. Consequently, the governing system is not repeated here.

\subsection{Finite-Volume Dual-Time Discretization}
\label{sec:fv_dual_time}

To numerically resolve the RF discharge dynamics, the governing equations must be appropriately discretized in both space and time. As a starting point for the finite-volume formulation, the dimensionless transport equations for all species and electron energy are grouped into a unified conservative form:
\begin{equation}
\frac{\partial \mathbf{Q}}{\partial t}
+
\nabla\cdot\mathbf{F}
=
\mathbf{S}.
\label{eq:compact_transport}
\end{equation}

In a general multi-species formulation, the state vector encompasses the number densities of all tracked species (e.g., electrons, ions, and metastables) along with the electron energy density:
\begin{equation}
\mathbf{Q}
=
\left(
n_e,
n_i,
n_*,
\varepsilon_e
\right)^T .
\end{equation}

Accordingly, the generalized flux vector $\mathbf{F}$ and the source vector $\mathbf{S}$ are assembled as
\begin{equation}
\mathbf{F}
=
\left(
\boldsymbol{\Gamma}_e,
\boldsymbol{\Gamma}_i,
\boldsymbol{\Gamma}_*,
\boldsymbol{\Gamma}_{\varepsilon}
\right)^T,
\end{equation}

\begin{equation}
\mathbf{S}
=
\left(
S_e,
S_i,
S_*,
S_{\varepsilon}
\right)^T,
\end{equation}
where $S_{\varepsilon} = \mathbf{J}_e\cdot\mathbf{E} - \sum_{j} \Delta\mathcal{E}_{j} S_{j}$ represents the net electron energy source term.

The computational domain is discretized into cell-centered control volumes. Integrating Eq.~\eqref{eq:compact_transport} over an arbitrary control volume $\Omega_i$ yields the semi-discrete finite-volume system:
\begin{equation}
\Omega_i
\frac{d\mathbf{Q}_i}{dt}
=
\mathbf{R}_i(\mathbf{Q},\phi),
\label{eq:semi_discrete_fv}
\end{equation}
where $\mathbf{Q}_i$ is the cell-averaged state and $\mathbf{R}_i$ is the spatial residual. The residual is evaluated as
\begin{equation}
\mathbf{R}_i
=
-
\sum_{f\in\partial\Omega_i}
A_f\mathbf{F}_f
+
\Omega_i\mathbf{S}_i, 
\label{eq:fv_residual}
\end{equation}
in which $A_f$ is the face area and $\mathbf{F}_f$ is the numerical flux projected along the outward unit normal vector $\mathbf{n}_f$ of face $f$. Explicitly, the projected face flux vector corresponding to the state variables is
\begin{equation}
\mathbf{F}_f
=
\left(
\boldsymbol{\Gamma}_{e,f}\cdot\mathbf{n}_f,
\boldsymbol{\Gamma}_{i,f}\cdot\mathbf{n}_f,
\boldsymbol{\Gamma}_{*,f}\cdot\mathbf{n}_f,
\boldsymbol{\Gamma}_{\varepsilon,f}\cdot\mathbf{n}_f
\right)^T.
\label{eq:face_flux}
\end{equation}

To advance the semi-discrete system in physical time, implicit backward differentiation formulas (BDF) are employed. The fully discrete nonlinear residual $\mathbf{G}_i$ at the new physical time level $n+1$ can be expressed in a unified BDF framework:
\begin{equation}
\mathbf{G}_i(\mathbf{Q}^{n+1})
=
\mathbf{R}_i(\mathbf{Q}^{n+1},\phi^{n+1})
-
\Omega_i
\frac{
\alpha_0\mathbf{Q}_i^{n+1}
+
\alpha_1\mathbf{Q}_i^{n}
+
\alpha_2\mathbf{Q}_i^{n-1}
}{\Delta t},
\label{eq:bdf_unified}
\end{equation}
where the coefficients are $(\alpha_0, \alpha_1, \alpha_2) = (1, -1, 0)$ for the first-order backward Euler scheme, and $(\alpha_0, \alpha_1, \alpha_2) = (3/2, -2, 1/2)$ for the second-order BDF2 scheme.

To solve the nonlinear system $\mathbf{G}_i=0$ at each physical time step, a pseudo-time relaxation approach is employed:
\begin{equation}
\Omega_i
\frac{\partial \mathbf{Q}_i}{\partial \tau}
=
\mathbf{G}_i(\mathbf{Q},\phi),
\label{eq:dual_time_equation}
\end{equation}
where the pseudo-time $\tau$ is a variable introduced exclusively for numerical relaxation; it does not correspond to any physical process. Its sole purpose is to asymptotically drive the implicit BDF residual to zero at each fixed physical time level. Consequently, the physical time step $\Delta t$ controls the temporal accuracy of the transient RF discharge dynamics, whereas the pseudo-time step $\Delta\tau$ represents the convergence rate of the nonlinear inner iterations.

To evaluate the numerical stability limit during the relaxation process, a local pseudo-time step restriction is first estimated for each cell $i$ based on a CFL-like condition:
\begin{equation}
\Delta\tau_i
=
\frac{{\rm CFL}_{\tau}}
{\lambda_i^{\rm tr}
+
\lambda_i^{\rm chem}
+
\lambda_i^{\rm P}},
\label{eq:local_pseudo_dt}
\end{equation}
where $\lambda_i^{\rm tr}$ represents the local spectral radius associated with drift-diffusion and energy transport, $\lambda_i^{\rm chem}$ estimates the stiffness of the chemical source terms, and $\lambda_i^{\rm P}$ accounts for the electrostatic (dielectric-relaxation) stiffness. However, because the electrostatic potential is governed by the elliptic Poisson equation---which introduces strong global spatial coupling---advancing the system with non-uniform local pseudo-time steps often disrupts this coupling and leads to severe numerical instability. Therefore, to ensure robust convergence, a global pseudo-time step is employed across the entire computational domain, taken as the minimum of the local limits: $\Delta\tau = \min_i (\Delta\tau_i)$. By employing this dual-time formulation, the physical temporal accuracy remains completely decoupled from the stability and convergence requirements of the inner iterations.

The convergence of the pseudo-time iterations is measured based on the full BDF residual $\mathbf{G}$. Let $m$ denote the inner-iteration index, $N_c$ the number of control volumes, and $N_q$ the number of transported variables in the present plasma model. The global fluid residual norm is defined as
\begin{equation}
\|\mathbf{G}^{(m)}\|_2
=
\left[
\frac{1}{N_cN_q}
\sum_{i=1}^{N_c}
\sum_{q=1}^{N_q}
\left(G_{i,q}^{(m)}\right)^2
\right]^{1/2}.
\label{eq:inner_res_norm}
\end{equation}

For each physical time step, the inner convergence is then monitored by the relative residual reduction
\begin{equation}
\rho_m
=
\frac{\|\mathbf{G}^{(m)}\|_2}{\|\mathbf{G}^{(0)}\|_2}.
\label{eq:inner_res_ratio}
\end{equation}

The nonlinear implicit update is considered converged when $\rho_m \le \epsilon_{\rm in}$, where $\epsilon_{\rm in}$ is the prescribed inner-iteration tolerance, or when the maximum number of inner iterations is reached. This criterion ensures that the pseudo-CFL number acts only as a relaxation parameter for solving the implicit BDF system and does not define the physical accuracy of the time integration. The residual of the Poisson equation is monitored separately because the electrostatic potential is obtained from an elliptic equation with a different scaling from the transported fluid variables.

\subsection{Implicit Relaxation and Update}
\label{sec:lusgs}

To drive the discrete nonlinear residual $\mathbf{G}_i=0$ to convergence at each physical time step, an implicit pseudo-time relaxation is employed. Let $m$ denote the current inner-iteration index, such that the state update is defined as $\Delta\mathbf{Q}_i = \mathbf{Q}_i^{m+1} - \mathbf{Q}_i^m$. Performing a first-order Taylor series expansion on the residual yields the unfactored linear system:
\begin{equation}
    \left[
    \left( \frac{\Omega_i}{\Delta\tau} + \frac{\alpha_0 \Omega_i}{\Delta t} \right)\mathbf{I}
    - \frac{\partial\mathbf{R}_i}{\partial\mathbf{Q}_i}
    \right] \Delta\mathbf{Q}_i
    - \sum_{j\in\mathcal{N}(i)}
    \frac{\partial\mathbf{R}_i}{\partial\mathbf{Q}_j} \Delta\mathbf{Q}_j
    = -\mathbf{G}_i(\mathbf{Q}^m, \phi^m),
    \label{eq:linear_system}
\end{equation}
where $\mathcal{N}(i)$ denotes the set of neighboring cells, and $\alpha_0$ is the temporal coefficient associated with the chosen BDF scheme. For steady-state or purely pseudo-transient simulations, the physical-time term proportional to $1/\Delta t$ vanishes. Therefore, the assembled linear system for cell $i$ can be abstracted as
\begin{equation}
    (D + L + U) \Delta\mathbf{Q} = \mathbf{b},
    \label{eq:dlu}
\end{equation}
in which $\mathbf{b}$ is the explicit residual vector, $D$ represents the block-diagonal matrix containing the temporal, transport, and chemistry terms, while $L$ and $U$ are the strictly lower and upper block-triangular matrices associated with the prescribed cell ordering. 

To avoid the computationally prohibitive direct inversion of the global matrix, the Lower-Upper Symmetric Gauss-Seidel (LU-SGS) method decomposes and factorizes the implicit operator. By extracting the diagonal matrix $D$, the operator is mathematically expanded as
\begin{align}
    D + L + U 
    &= D \left( \mathbf{I} + D^{-1}L + D^{-1}U \right) \nonumber \\
    &= D \left( \mathbf{I} + D^{-1}L \right) \left( \mathbf{I} + D^{-1}U \right) - L D^{-1} U.
    \label{eq:lusgs_expansion}
\end{align}

Since its contribution scales with the state increment $\Delta\mathbf{Q}$, the cross-coupling term $L D^{-1} U$ naturally vanishes as the pseudo-transient state converges. Therefore, it can be safely neglected during the relaxation process without affecting the final converged solution, leading to the following approximate factorization:
\begin{align}
    D + L + U 
    &\approx D \left( \mathbf{I} + D^{-1}L \right) \left( \mathbf{I} + D^{-1}U \right) \nonumber \\
    &= (D + L) D^{-1} (D + U)
    \label{eq:lusgs_approx}
\end{align}

Substituting this approximate factorization back into the global linear system yields the standard LU-SGS formulation:
\begin{equation}
    (D + L) D^{-1} (D + U) \Delta\mathbf{Q} = \mathbf{b}.
    \label{eq:lusgs_fact}
\end{equation}

This approximate factorization is efficiently inverted through a two-step symmetric sweep process. The forward sweep computes an intermediate update $\Delta\mathbf{Q}^*$ by solving the lower-triangular system:
\begin{equation}
    (D + L) \Delta\mathbf{Q}^{*} = \mathbf{b}.
    \label{eq:forward_global}
\end{equation}

At the discrete cell level, this global operation translates to a local, cell-by-cell update:
\begin{equation}
    D_i \Delta\mathbf{Q}_i^{*}
    = \mathbf{b}_i - \sum_{j<i} L_{ij} \Delta\mathbf{Q}_j^{*},
    \label{eq:forward_local}
\end{equation}
where $j<i$ designates adjacent cells that have already been updated during the current forward traversal. 

Subsequently, the backward sweep determines the final state increment $\Delta\mathbf{Q}$ by solving the upper-triangular system:
\begin{equation}
    D^{-1} (D + U) \Delta\mathbf{Q} = \Delta\mathbf{Q}^{*}.
    \label{eq:backward_global}
\end{equation}

Multiplying by the diagonal block and rearranging, the local update during the reverse traversal becomes:
\begin{equation}
    D_i \Delta\mathbf{Q}_i
    = D_i \Delta\mathbf{Q}_i^{*} - \sum_{j>i} U_{ij} \Delta\mathbf{Q}_j,
    \label{eq:backward_local}
\end{equation}
where $j>i$ denotes neighbors that have already been updated in the backward sweep. In the present implementation, $D_i$ is a compact local block matrix that is inverted directly at minimal computational cost.
The robustness of the LU-SGS method relies on the construction of the local diagonal block $D_i$, which incorporates pseudo-time, physical-time, spatial transport, and chemical source contributions:
\begin{equation}
    D_i =
    \left( \frac{\Omega_i}{\Delta\tau} + \frac{\alpha_0 \Omega_i}{\Delta t} \right) \mathbf{I}
    + D_i^{\rm tr} 
    - \Omega_i \mathbf{J}_i^{\rm chem}.
\end{equation}

To circumvent the prohibitive storage requirements and computational cost of assembling the exact spatial flux Jacobians, a low-storage LU-SGS approximation is adopted. The transport Jacobian is rendered strongly diagonally dominant and approximated using the local maximum spectral radii:
\begin{equation}
    D_i^{\rm tr}
    \approx \Omega_i \operatorname{diag}(\lambda_{e,i}, \lambda_{i,i}, \lambda_{*,i}, \lambda_{\varepsilon,i}).
\end{equation}

For drift-diffusion processes on general meshes, these spectral radii are evaluated over all boundary faces $f \in \partial\Omega_i$ as
\begin{align}
    \lambda_{s,i} &\approx \sum_{f\in\partial\Omega_i}
    \frac{A_f}{|\Omega_i|}
    \left( \frac{1}{2}\mu_s |\mathbf{E}_f| + c_D\frac{D_s}{|\Delta \mathbf{r}_f|} \right), \quad (s = e, i, *) ,\\
    \lambda_{\varepsilon,i} &\approx \sum_{f\in\partial\Omega_i}
    \frac{A_f}{|\Omega_i|}
    \left( \frac{5}{6}\mu_e |\mathbf{E}_f| + \frac{5}{3}c_D\frac{D_e}{|\Delta \mathbf{r}_f|} \right),
\end{align}
where $c_D$ is a numerical diffusion scaling constant and $|\Delta \mathbf{r}_f|$ is the characteristic distance between neighboring cell centers. Note that for uncharged metastables, the drift mobility $\mu_*$ is strictly zero. 

The chemistry Jacobian $\mathbf{J}_i^{\rm chem} = \partial\mathbf{S}_i/\partial\mathbf{Q}_i$ provides the necessary implicit coupling for the reactive source terms. In a general multi-species framework, the net source terms are formulated as combinations of reaction rate coefficients and explicit species number densities. 

While the rate coefficients are strongly dependent on the mean electron energy, evaluating their exact analytical derivatives can be mathematically cumbersome. To simplify the construction of the implicit operator, the rate coefficients are treated as locally frozen parameters during the linearization. By neglecting their variations with respect to the state variables, the Jacobian entries are efficiently approximated by differentiating only the explicit species concentrations. Consequently, the exact implicit dependence of the chemistry source terms on the electron energy is dropped in the preconditioner block. 

Furthermore, to guarantee matrix invertibility and numerical stability under extremely stiff chemical transients, a row-sum damping technique is applied to the local diagonal block $D_i$. Specifically, for each state variable, the absolute sum of the off-diagonal elements in the corresponding row of the chemistry Jacobian is multiplied by a user-controlled damping coefficient $\eta_{\rm chem} \ge 0$, and this resulting value is explicitly added to the main diagonal entry.

Finally, to ensure physical realizability---specifically, strictly positive species densities and electron energy---the computed LU-SGS correction is applied with a dynamic relaxation limiter. The local state vector is updated as
\begin{equation}
    \mathbf{Q}_i^{m+1} = \mathbf{Q}_i^m + \alpha \Delta\mathbf{Q}_i, \qquad 0 < \alpha \le 1,
\end{equation}
where the under-relaxation factor $\alpha$ is dynamically determined to prevent non-physical negative values during severe numerical transients.

\subsection{Semi-Implicit Poisson Update}

Solving Poisson's equation decoupled from the particle transport often imposes a severe dielectric-relaxation time step limit, leading to weak electrostatic coupling and numerical instability. To alleviate this stiffness and ensure robust convergence, we employ a semi-implicit correction, following the approaches established in \cite{kushner2009hybrid,li2025fast,ventzek1993two}. This method approximates the fast electron number density response to potential variations over the pseudo-time step $\Delta\tau$.

To construct this semi-implicit operator, the implicit treatment is exclusively applied to the electron drift term. The electron diffusion is evaluated explicitly, while the chemical reaction source terms are neglected during this predictor step. Consequently, starting from the electron continuity equation, the semi-implicit density update is approximated by:
\begin{equation}
    n_e^{m+1}
    \approx n_e^{m}
    - \Delta\tau \nabla\cdot\boldsymbol{\Gamma}_{e}^{m+1},
\end{equation}
where $m$ is the pseudo-time inner iteration index. By keeping the electric field implicit while evaluating the mobility and electron number density at the known iterate step $m$, the implicit drift flux is given as
\begin{equation}
    \boldsymbol{\Gamma}_{e}^{m+1}
    \approx -\mu_e n_e^m \mathbf{E}^{m+1} 
    = \mu_e n_e^m \nabla\phi^{m+1} .
\end{equation}

Substituting this equation into Poisson's equation yields the modified semi-implicit operator
\begin{equation}
    -\nabla\cdot\left(\epsilon_0\nabla\phi^{m+1}\right)
    - \nabla\cdot\left(e\mu_e n_e^m\Delta\tau\nabla\phi^{m+1}\right)
    = e(n_i^m-n_e^m)+\mathcal{D}_e^m,
    \label{eq:semi_poisson}
\end{equation}
where $\mathcal{D}_e^m$ represents the explicit electron diffusion contributions, given by
\begin{equation}
    \mathcal{D}_e^m
    =
    -e \Delta\tau \nabla\cdot\left(D_e \nabla n_e^m\right).
    \label{eq:explicit_D}
\end{equation}

Through this formulation, the original Laplacian operator is augmented by a positive-definite drift-response term, proportional to $e\mu_e n_e^m\Delta\tau$, which mathematically strengthens the diagonal dominance of the global matrix, thereby effectively damping the dielectric-relaxation stiffness and accelerating the electrostatic coupling. Upon finite-volume spatial discretization, Eq.~\eqref{eq:semi_poisson} produces a sparse linear system. To maintain accuracy on distorted elements, the face-normal potential gradients are discretized incorporating non-orthogonal cross-diffusion corrections.

\subsection{Boundary Conditions}
\label{sec:boundary_conditions}

To complete the physical model and close the semi-discrete finite-volume system, physically consistent boundary conditions must be specified at the solid electrode surfaces. In the present formulation, the electrostatic potential is imposed via Dirichlet boundary conditions, while the transport equations for charged particles and electron energy are closed through wall-flux boundary conditions.

For an RF capacitively coupled discharge, the powered electrode is driven by a prescribed sinusoidal voltage:
\begin{equation}
    \phi_{\rm p}(t) = V_{\rm RF}\sin\left(2\pi f_{\rm RF}t+\varphi_0\right),
    \label{eq:rf_potential_bc}
\end{equation}
where $V_{\rm RF}$ is the peak voltage amplitude, $f_{\rm RF}$ is the driving frequency, and $\varphi_0$ is the phase shift. The grounded electrode is fixed at a reference potential $\phi_{\rm g} = 0$.

Driven predominantly by the strong sheath electric field, ions are accelerated toward the wall. The theoretical drift-driven ion flux at the boundary is given by
\begin{equation}
    \boldsymbol{\Gamma}_i\cdot\mathbf{n}_w = \mu_i n_i \mathbf{E}\cdot\mathbf{n}_w,
    \label{eq:ion_wall_bc_theory}
\end{equation}
where $\mathbf{n}_w$ is the outward unit normal vector pointing from the plasma domain to the wall. In the actual numerical implementation, to prevent unphysical inward ion drift during highly transient iterative steps, a pure upwind condition is enforced by limiting this flux to non-negative values (i.e., $\max(\boldsymbol{\Gamma}_i\cdot\mathbf{n}_w, 0)$).

For the electron and electron energy fluxes, the physical treatment at the wall depends heavily on the thermal assumptions. In this work, two distinct sets of flux boundary conditions are implemented to accommodate different benchmarking requirements:

\paragraph{1. Energy-Independent Flux (Prescribed Wall Temperature)}

The first set assumes an energy-independent electron flux, typically associated with a prescribed constant electron temperature at the wall. The macroscopic electron loss is modeled using a constant surface recombination velocity:
\begin{equation}
    \boldsymbol{\Gamma}_e\cdot\mathbf{n}_w = k_s n_e - \gamma (\boldsymbol{\Gamma}_i\cdot\mathbf{n}_w),
    \label{eq:electron_wall_bc_indep}
\end{equation}
where $k_s$ is the effective electron surface recombination coefficient (assuming a unity sticking coefficient) and $\gamma$ is the secondary electron emission (SEE) coefficient. Because the flux does not depend on the local energy, the electron energy equation must be closed by explicitly specifying a Dirichlet boundary condition based on the prescribed wall temperature $T_{e,w}$:
\begin{equation}
    \varepsilon_{e,w} = \frac{3}{2} n_{e,w} k_B T_{e,w}.
    \label{eq:wall_energy_bc}
\end{equation}

\paragraph{2. Energy-Dependent Flux (Zero-Gradient Temperature Extrapolation)}

The second set introduces electron energy dependence derived from kinetic theory, which assumes the electron temperature at the boundary is extrapolated from the adjacent interior cell (i.e., a zero-gradient condition for $T_e$). The macroscopic loss rate is governed by the thermal flux of a half-Maxwellian distribution:
\begin{equation}
    \boldsymbol{\Gamma}_e\cdot\mathbf{n}_w = \frac{1}{4} n_e \left( \frac{8 k_B T_e}{\pi m_e} \right)^{1/2} - \gamma \boldsymbol{\Gamma}_i\cdot\mathbf{n}_w,
    \label{eq:electron_wall_bc_dep}
\end{equation}
where $T_e$ is the locally extrapolated electron temperature, and $m_e$ is the electron mass. Consistent with this particle flux, the transported electrons carry their respective mean thermal energies to the wall. The corresponding energy boundary flux is evaluated as:
\begin{equation}
    \boldsymbol{\Gamma}_{\varepsilon}\cdot\mathbf{n}_w = \frac{5}{3} \bar{\varepsilon}_e \left[ \frac{1}{4} n_e \left( \frac{8 k_B T_e}{\pi m_e} \right)^{1/2} - \gamma \boldsymbol{\Gamma}_i\cdot\mathbf{n}_w \right].
    \label{eq:energy_wall_flux_bc}
\end{equation}

\begin{figure*}[!h]
    \centering
    \includegraphics[width=0.95\textwidth]{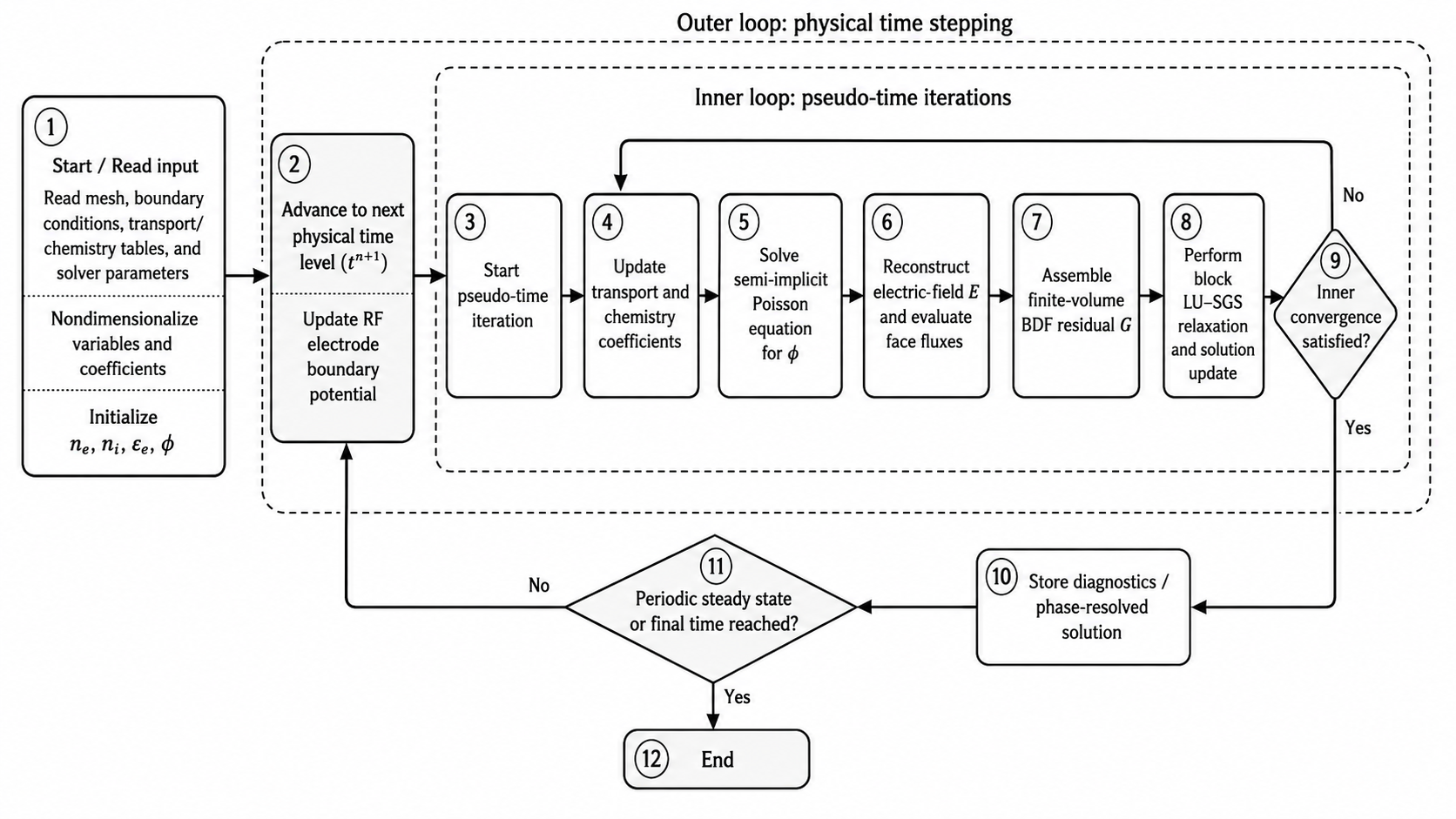}
    \caption{Compact algorithmic flow of the dual-time implicit finite-volume RF CCP solver.}
    \label{fig:algorithm_flow}
\end{figure*}

The overall solution procedure is summarized in Fig.~\ref{fig:algorithm_flow}. At each physical time level, the backward-difference residual is iteratively driven to zero in pseudo-time. Within each inner iteration, the electrostatic potential is updated from the semi-implicit Poisson equation, after which the face fluxes, chemistry source terms, and finite-volume residuals are evaluated. The resulting linearized system is relaxed by block LU-SGS sweeps, followed by a positivity-preserving update of the transported variables.

\section{Numerical Setup and Results}

\subsection{Quasi-One-dimensional RF CCP benchmark}
\subsubsection{Numerical Setup}

\begin{figure}[h]
    \centering
    \includegraphics[width=0.8\textwidth]{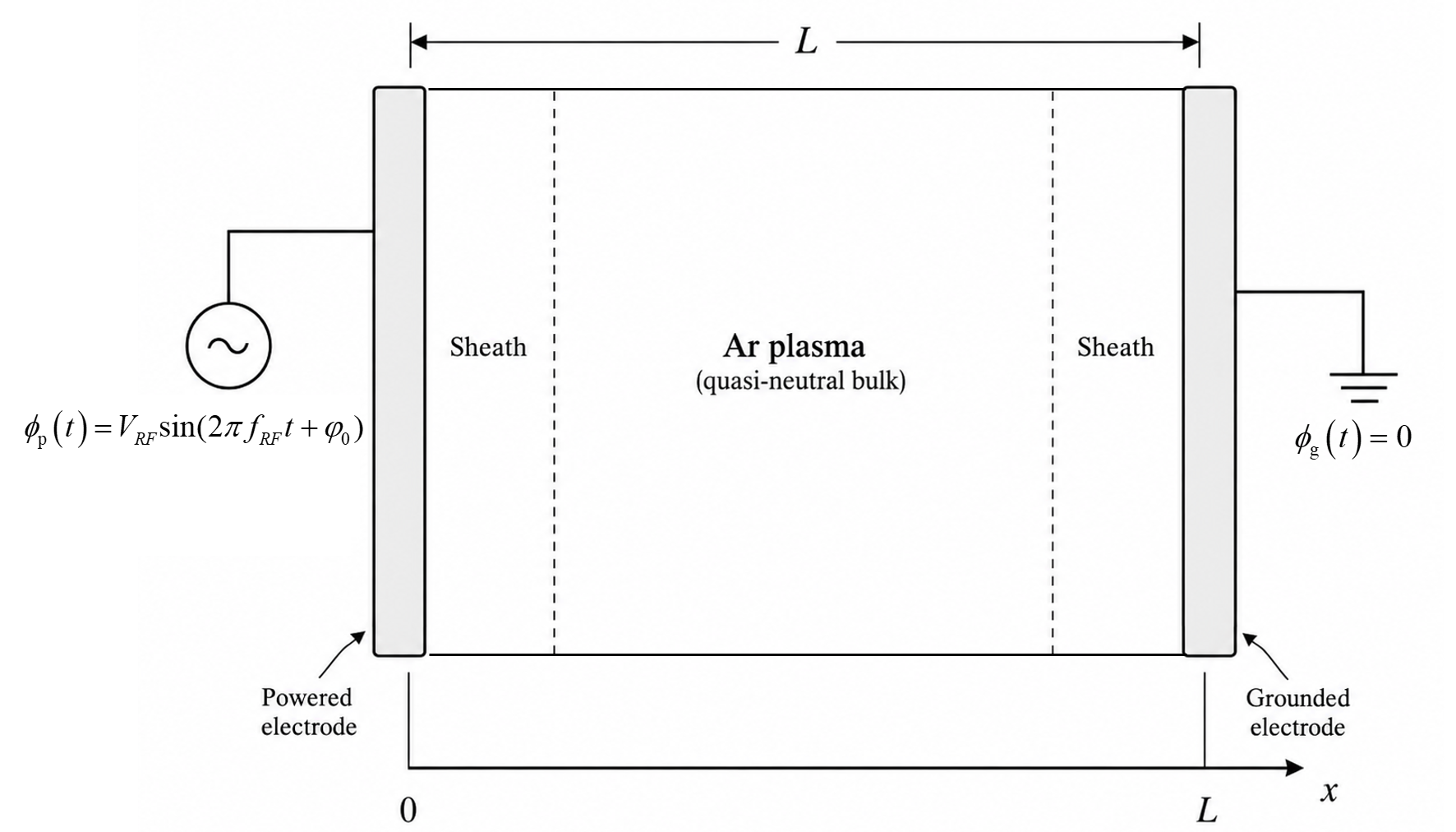}
    \caption{Schematic diagram of the one-dimensional parallel-plate radio-frequency argon capacitively coupled plasma (CCP) benchmark.}
    \label{fig:glow_discharge}
\end{figure}

The numerical tests are based on a reference benchmark configuration of a one-dimensional, parallel-plate radio-frequency argon capacitively coupled plasma, as schematically illustrated in Fig.~\ref{fig:glow_discharge}. 

Correspondingly, the primary physical and geometric parameters governing the discharge are summarized in Table~\ref{tab:physical_parameters}.

\begin{table}[h]
    \centering
    \caption{Physical parameters for the RF argon CCP benchmark.}
    \label{tab:physical_parameters}
    \begin{tabular}{lll}
        \toprule
        \textbf{Parameter} & \textbf{Symbol} & \textbf{Value} \\
        \midrule
        Interelectrode distance & $L_x$ & 2.54~cm \\
        Gas pressure & $p$ & 1~Torr \\
        Gas temperature & $T_g$ & 300~K \\
        Neutral argon density & $N_g$ & $3.22 \times 10^{22}~\mathrm{m^{-3}}$ \\
        Driving voltage amplitude & $V_0$ & 100~V \\
        Driving frequency & $f_{\mathrm{rf}}$ & 13.56~MHz \\
        Surface recombination  velocity & $k_s$ & $1.19 \times 10^5~\mathrm{m/s}$ \\
        Secondary emission coefficient & $\gamma$ & 0.01 \\
        Wall electron temperature & $T_{e,w}$ & 0.5~eV \\
        \bottomrule
    \end{tabular}
\end{table}

A sinusoidal voltage, $\phi(0,t) = V_0\sin(2\pi f_{\mathrm{rf}}t)$, is applied to the powered electrode at $x=0$, while the opposite electrode at $x=L_x$ is grounded ($\phi(L_x,t) = 0$). Although the reference discharge is strictly one-dimensional, the present finite-volume solver is executed on a two-dimensional computational domain. Consequently, the benchmark is represented by a narrow rectangular computational domain with a length of $L_x = 0.0254~\mathrm{m}$ in the $x$-direction and a height of $L_y = 0.005~\mathrm{m}$ in the $y$-direction. The $x$-direction is normal to the electrodes, whereas the $y$-direction is treated as homogeneous. The horizontal grid employs a nonuniform node distribution to cluster cells near both electrodes, ensuring sufficient resolution for the sharp sheath gradients. The node coordinates $x_i$ are defined as:
\begin{equation}
    x_i = \frac{L_x}{2} \left[ \frac{i-1}{(N_p-1)/2} \right]^2, \qquad i=1,\ldots,\frac{N_p-1}{2}+1,
\end{equation}
\begin{equation}
    x_{N_p-i+1} = L_x-x_i, \qquad i=1,\ldots,\frac{N_p-1}{2}+1,
\end{equation}
where $N_p=91$ is the total number of nodes, yielding 90 cells in the interelectrode direction. In the transverse ($y$) direction, only one uniform cell is used.

To ensure the 2D solution accurately reflects the 1D physics, zero-gradient symmetry boundary conditions are strictly enforced at the upper and lower boundaries of the strip:
\begin{equation}
    \boldsymbol{\Gamma}_e \cdot \mathbf{n} = 0, \quad
    \boldsymbol{\Gamma}_i \cdot \mathbf{n} = 0, \quad
    \boldsymbol{\Gamma}_{\varepsilon} \cdot \mathbf{n} = 0, \quad
    \nabla\phi \cdot \mathbf{n} = 0.
\end{equation}

The simulation utilizes a reduced argon chemistry model encompassing two primary electron-impact reactions. The first reaction is direct ionization:
\begin{equation}
    e + \mathrm{Ar} \rightarrow 2e + \mathrm{Ar}^{+},
\end{equation}
which generates a new electron-ion pair, acting as the fundamental particle source that sustains the discharge. The second reaction is excitation:
\begin{equation}
    e + \mathrm{Ar} \rightarrow e + \mathrm{Ar}^{*}.
\end{equation}

Although excitation does not alter the total number of charged particles, it consumes a substantial amount of electron kinetic energy, thereby constituting a primary mechanism for electron-energy loss in the system. The corresponding macroscopic reaction rates are expressed as:
\begin{equation}
    R_{\mathrm{ion}} = N_g n_e k_{\mathrm{ion}}(\bar{\varepsilon}_e),
\end{equation}
\begin{equation}
 R_{\mathrm{exc}} = N_g n_e k_{\mathrm{exc}}(\bar{\varepsilon}_e),
\end{equation}
where the rate coefficients $k_{\mathrm{ion}}$ and $k_{\mathrm{exc}}$ are evaluated as functions of the mean electron energy, $\bar{\varepsilon}_e = \varepsilon_e / (n_e e)$, given in units of eV. 

\begin{figure}[h]
    \centering
    \includegraphics[width=0.7\textwidth]{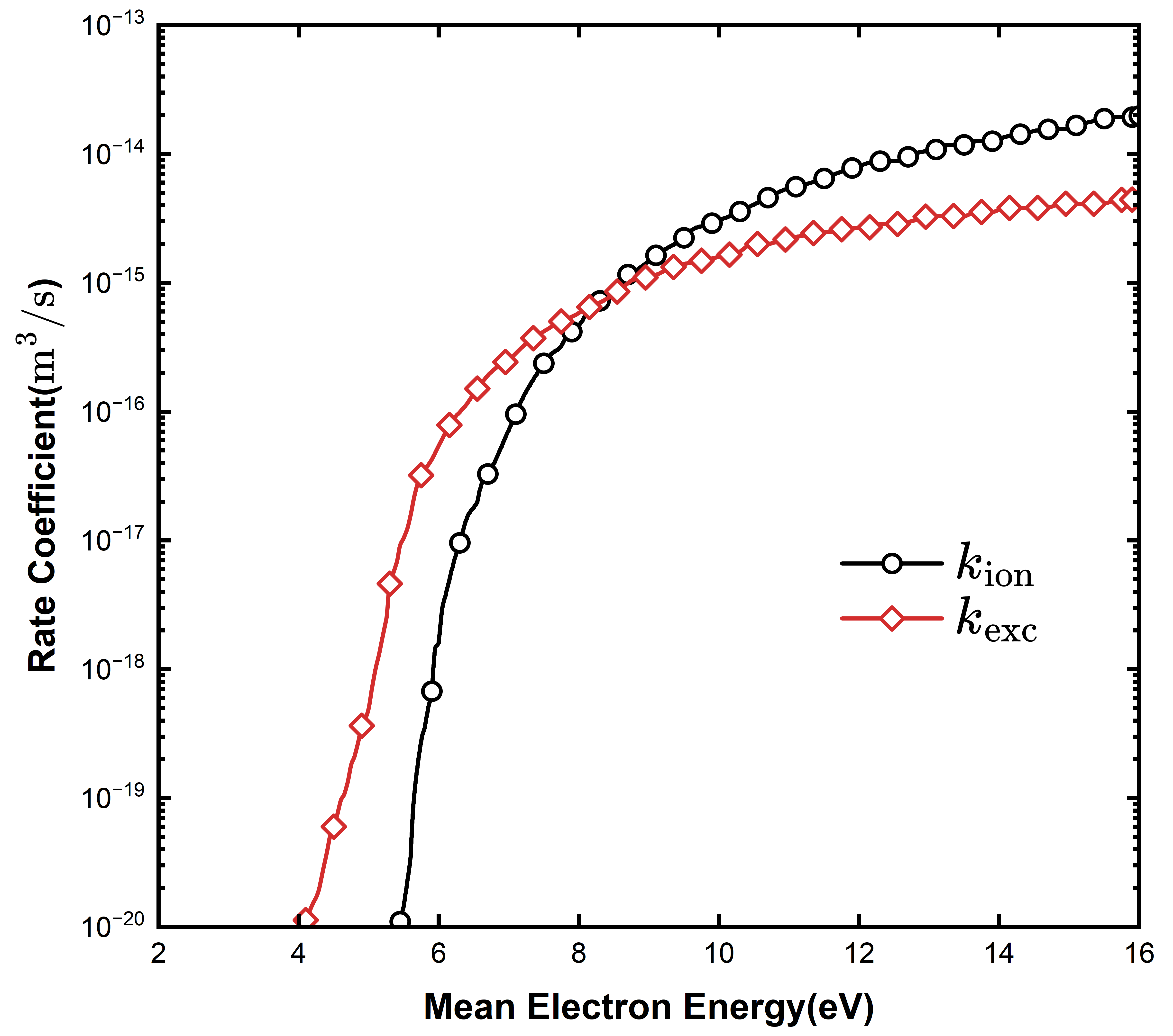}
    \caption{Rate coefficients for direct ionization and excitation of argon as a function of mean electron energy.}
    \label{fig:rate_coefficients}
\end{figure}

In standard physical modeling, macroscopic rate coefficients are typically derived by solving the two-term approximation of the electron Boltzmann equation using tools such as BOLSIG+~\cite{hagelaar2016brief,hagelaar2016coulomb}. This conventional approach involves calculating the non-Maxwellian electron energy distribution function (EEDF) and subsequently integrating the respective energy-dependent collision cross-sections over the computed EEDF~\cite{pancheshnyi2012lxcat}. However, to ensure a rigorous and unambiguous code verification against the one-dimensional reference calculation, the present simulations adopt the exact mean-energy-dependent rate tables utilized in the reference benchmark~\cite{lymberopoulos1993fluid}. The variation of these rate coefficients with mean electron energy is depicted in Fig.~\ref{fig:rate_coefficients}.

Since no metastable continuity equation is solved in this reduced model, ionization acts as the sole source term for both the electron and ion continuity equations ($S_e = S_i = R_{\mathrm{ion}}$). Excitation contributes exclusively to the inelastic electron-energy loss. Therefore, the total energy source term is formulated as:
\begin{equation}
    S_{\varepsilon} = -\Delta\varepsilon_{\mathrm{ion}} R_{\mathrm{ion}} -\Delta\varepsilon_{\mathrm{exc}} R_{\mathrm{exc}},
\end{equation}
with threshold energies $\Delta\varepsilon_{\mathrm{ion}} = 15.7~\mathrm{eV}$ and $\Delta\varepsilon_{\mathrm{exc}} = 11.56~\mathrm{eV}$. To maintain rigorous consistency with the 1D reference calculation, the rate coefficients are interpolated directly from the reference benchmark's mean-energy-dependent tables. Logarithmic interpolation is applied within the positive-rate region to preserve strict numerical positivity across the wide dynamic range of the cross sections.

The drift-diffusion transport of electrons and ions is governed by their respective mobilities ($\mu$) and diffusivities ($D$). In this benchmark, these transport coefficients are assumed to be inversely proportional to the neutral gas density $N_g$. The constant products of the neutral density and the transport coefficients utilized in the present simulation are summarized in Table~\ref{tab:transport_coefficients}.
\begin{table}[h]
    \centering
    \caption{Reduced transport coefficients for electrons and argon ions.}
    \label{tab:transport_coefficients}
    \begin{tabular}{llll}
        \toprule
        \textbf{Name} & \textbf{Symbol} & \textbf{Unit} & \textbf{Value} \\
        \midrule
        Electron diffusivity product & $N_g D_e$   & $\mathrm{m^{-1}\,s^{-1}}$           & $3.86 \times 10^{24}$ \\
        Electron mobility product    & $N_g \mu_e$ & $\mathrm{V^{-1}\,m^{-1}\,s^{-1}}$ & $9.66 \times 10^{23}$ \\
        Ion diffusivity product      & $N_g D_i$   & $\mathrm{m^{-1}\,s^{-1}}$           & $2.07 \times 10^{20}$ \\
        Ion mobility product         & $N_g \mu_i$ & $\mathrm{V^{-1}\,m^{-1}\,s^{-1}}$ & $4.65 \times 10^{21}$ \\
        \bottomrule
    \end{tabular}
\end{table}

At the electrode surfaces, the electrostatic potential is governed by the aforementioned RF and ground conditions. For charged-particle transport, an energy-independent wall-flux formulation is applied. Following the outward-pointing normal vector $\mathbf{n}_w$, the electron wall flux incorporates both thermal loss and secondary electron emission induced by ion impact:
\begin{equation}
    \boldsymbol{\Gamma}_e \cdot \mathbf{n}_w = k_s n_e - \gamma (\boldsymbol{\Gamma}_i \cdot \mathbf{n}_w).
\end{equation}

The boundary condition for the electron-energy equation is dictated by the prescribed wall electron temperature $T_{e,w}$, which explicitly defines the energy flux leaving the domain at the physical boundaries.

\subsubsection{Numerical Result}

For low-temperature RF plasma simulations, temporal accuracy is strongly affected by the physical time-step resolution. In the present dual-time implicit framework, the BDF residual at each physical time level is solved by pseudo-time iterations, so that the physical time step is not constrained by explicit stability limits. Instead, it serves as a prescribed temporal discretization parameter that must be sufficiently small to resolve the RF-periodic discharge dynamics. This requirement is especially relevant to capacitively coupled plasmas, where the sheath motion, electric-field variation, charged-particle transport, and ionization source terms exhibit pronounced nonlinear changes within one RF period. Consequently, stable computations obtained with large physical time steps may still lead to inaccurate periodic solutions if the RF cycle and sheath dynamics are under-resolved. A physical time-step verification is therefore performed to confirm that the selected time step yields adequate temporal accuracy for the subsequent analysis.

\begin{figure}[h]
    \centering
    \begin{subfigure}[t]{0.49\textwidth}
        \centering
        \includegraphics[width=\textwidth]{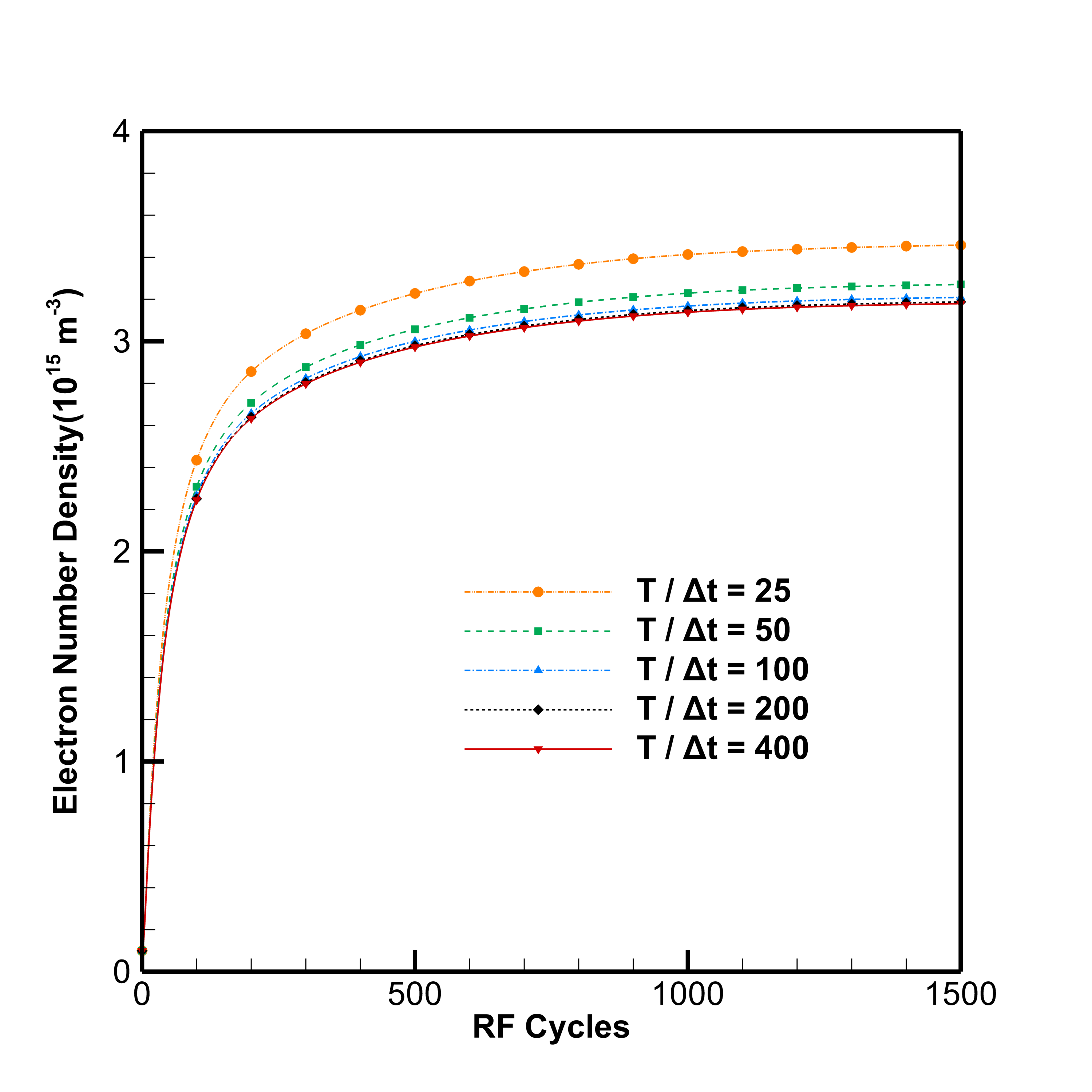}
        \caption{Electron number density.}
        \label{fig:dt_independence_ne}
    \end{subfigure}
    \hfill
    \begin{subfigure}[t]{0.49\textwidth}
        \centering
        \includegraphics[width=\textwidth]{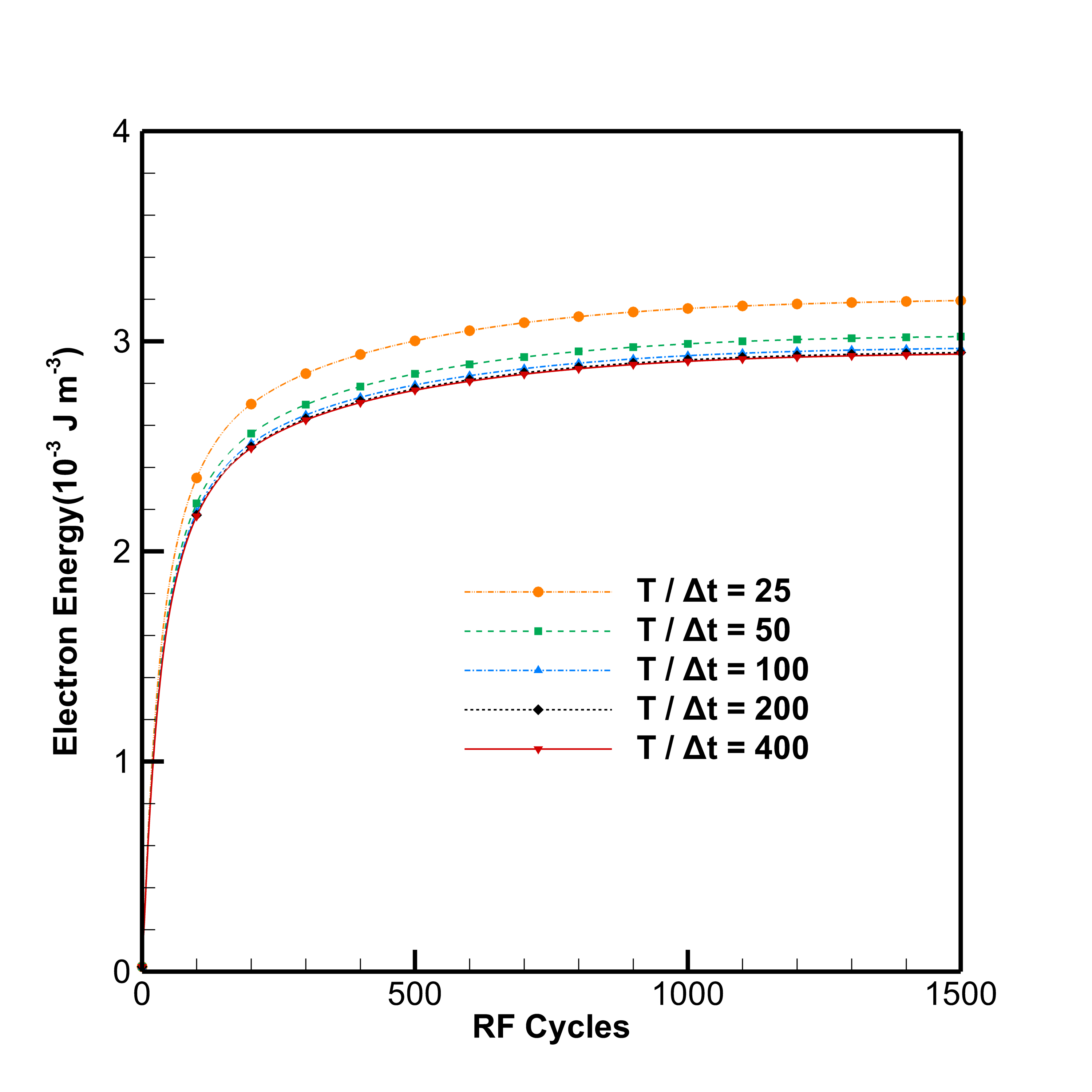}
        \caption{Electron energy density.}
        \label{fig:dt_independence_ee}
    \end{subfigure}
    \caption{Physical time-step verification based on the cycle-to-cycle evolution of spatially averaged plasma quantities sampled at the beginning of each RF period: 
    (a) electron number density and (b) electron energy density.}
    \label{fig:dt_independence}
\end{figure}

Figure~\ref{fig:dt_independence} shows the cycle-to-cycle evolution of the spatially averaged electron number density and electron energy density sampled at the beginning of each RF cycle for $T/\Delta t = 25$, $50$, $100$, $200$, and $400$, where $T$ denotes the RF period. All cases are computed based on the same spatial mesh, BDF2 time integration, and 100 pseudo-time inner iterations per physical time step. In the pseudo-time iteration, a pseudo-CFL number of $10000$ is used, corresponding to a pseudo-time step of approximately $7.64\times10^{-10} \mathrm{s}$.
  As the physical time step is refined, both averaged quantities gradually approach a common asymptotic behavior. The coarsest case, $T/\Delta t = 25$, clearly overpredicts the converged density and energy levels, indicating that the RF-periodic dynamics are insufficiently resolved despite the numerical stability of the dual-time iteration. In contrast, the results obtained with $T/\Delta t = 100$, $200$, and $400$ are much closer to each other, suggesting that the temporal discretization error has been significantly reduced.

 The quantitative differences are summarized in Table~\ref{tab:dt_verification}.

\begin{table}[h]
    \centering
    \caption{Physical time-step verification based on cycle-start spatial averages at the 1500th RF cycle. Relative errors are computed using the $T/\Delta t=400$ result as the reference.}
    \label{tab:dt_verification}
    \begin{tabular}{ccccc}
        \toprule
        $T/\Delta t$ &
        \begin{tabular}{c}
            $\langle n_e \rangle$ \\
            ($10^{15}~\mathrm{m^{-3}}$)
        \end{tabular} &
        \begin{tabular}{c}
            $\langle \varepsilon_e \rangle$ \\
            ($10^{-3}~\mathrm{J\,m^{-3}}$)
        \end{tabular} &
        \begin{tabular}{c}
            $E_{n_e}$ \\
            (\%)
        \end{tabular} &
        \begin{tabular}{c}
            $E_{\varepsilon_e}$ \\
            (\%)
        \end{tabular} \\
        \midrule
        25  & 3.45780 & 3.19394 & 8.76 & 8.67 \\
        50  & 3.27036 & 3.02250 & 2.86 & 2.84 \\
        100 & 3.20901 & 2.96630 & 0.93 & 0.93 \\
        200 & 3.18727 & 2.94634 & 0.25 & 0.25 \\
        400 & 3.17930 & 2.93904 & --   & --   \\
        \bottomrule
    \end{tabular}
\end{table}

The finest case, $T/\Delta t = 400$, is taken as the reference solution, and the relative errors are evaluated after 1500 RF cycles. As $T/\Delta t$ increases from $25$ to $100$, the relative errors of both the spatially averaged electron number density and electron energy density decrease from approximately $9\%$ to below $1\%$. Further refinement to $T/\Delta t = 200$ reduces the error to approximately $0.25\% $, confirming the convergence of the physical time discretization. Considering both temporal accuracy and computational cost, $T/\Delta t = 100$ is adopted in the following simulations unless otherwise stated.

This result also highlights the advantage of the present dual-time implicit formulation for RF plasma simulations. In conventional physical-time marching approaches, particularly those based on first-order backward Euler integration, the physical time step is often restricted by both numerical robustness and temporal accuracy. As a result, several thousand time steps per RF cycle may be required to obtain a sufficiently resolved periodic response, especially in the presence of rapid sheath motion, strong electric-field gradients, and stiff plasma chemistry. In contrast, the present formulation solves an implicit BDF residual at each physical time level through pseudo-time iterations, allowing the physical time step to be selected primarily according to the desired temporal accuracy rather than by a restrictive stability condition. The verification above demonstrates that $T/\Delta t = 100$ already provides a sufficiently accurate periodic solution for the present benchmark, offering an effective compromise between accuracy and computational efficiency.

\begin{figure}[h]
    \centering
    \includegraphics[width=0.72\textwidth]{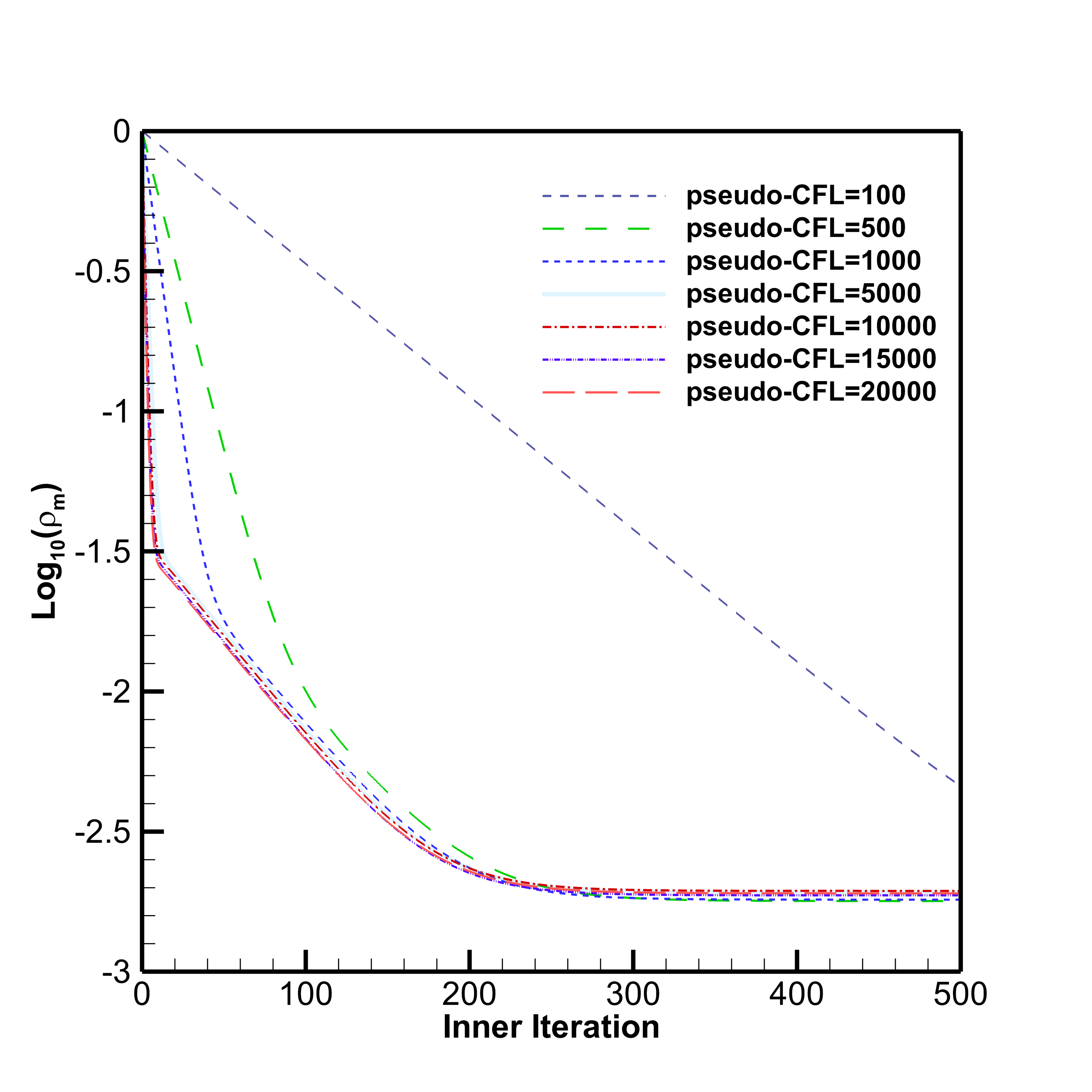}
    \caption{Pseudo-time convergence histories for different pseudo-CFL numbers based on the relative BDF residual.}
    \label{fig:cfl_convergence}
\end{figure}

The computational efficiency of the present dual-time implicit scheme is also affected by the convergence of the pseudo-time relaxation used to solve the implicit BDF residual at each physical time level. In contrast to the physical time step, which determines the temporal resolution of the RF-periodic discharge, the pseudo-time step is an algorithmic parameter that controls the relaxation rate of the nonlinear implicit residual. In the present test, the pseudo-CFL number is varied from $100$ to $20000$, corresponding to pseudo-time steps ranging from approximately $7.64\times10^{-12}$ to $1.53\times10^{-9}$. Therefore, the tested pseudo-time steps span about 2.3 orders of magnitude, providing a sufficiently broad range for assessing the effect of pseudo-time relaxation on inner-iteration convergence. Figure~\ref{fig:cfl_convergence} compares the inner-iteration histories of the relative BDF residual for different pseudo-CFL numbers.

A clear dependence of the convergence rate on the pseudo-CFL number can be observed. For a small pseudo-CFL number, such as pseudo-CFL number being 100, the residual decreases slowly and does not reach the common asymptotic level within 500 inner iterations. This behavior is consistent with the strong pseudo-time diagonal damping introduced by a small pseudo-time step, which limits the update magnitude during each relaxation iteration. Increasing the pseudo-CFL number to 500 and 1000 substantially accelerates the residual reduction, although the convergence is still noticeably slower than that obtained with larger pseudo-CFL numbers. When pseudo-CFL number is larger than 5000, the residual histories become very close to one another: all cases exhibit a rapid initial decrease during the first few iterations, followed by a slower asymptotic decay, and eventually approach a similar residual plateau of approximately $\log_{10}(\rho_m)\approx -2.7$ after about 200 inner iterations. Further increasing the pseudo-CFL number beyond 5000 provides little additional improvement, indicating that the convergence rate is no longer mainly limited by the pseudo-time diagonal term. Instead, the asymptotic behavior is likely governed by the approximate LU-SGS factorization, the nonlinear source-term coupling, and the interaction between the fluid equations and the semi-implicit Poisson update.

\begin{figure}[htbp]
    \centering
    \begin{minipage}{0.48\textwidth}
        \centering
        \includegraphics[width=\linewidth]{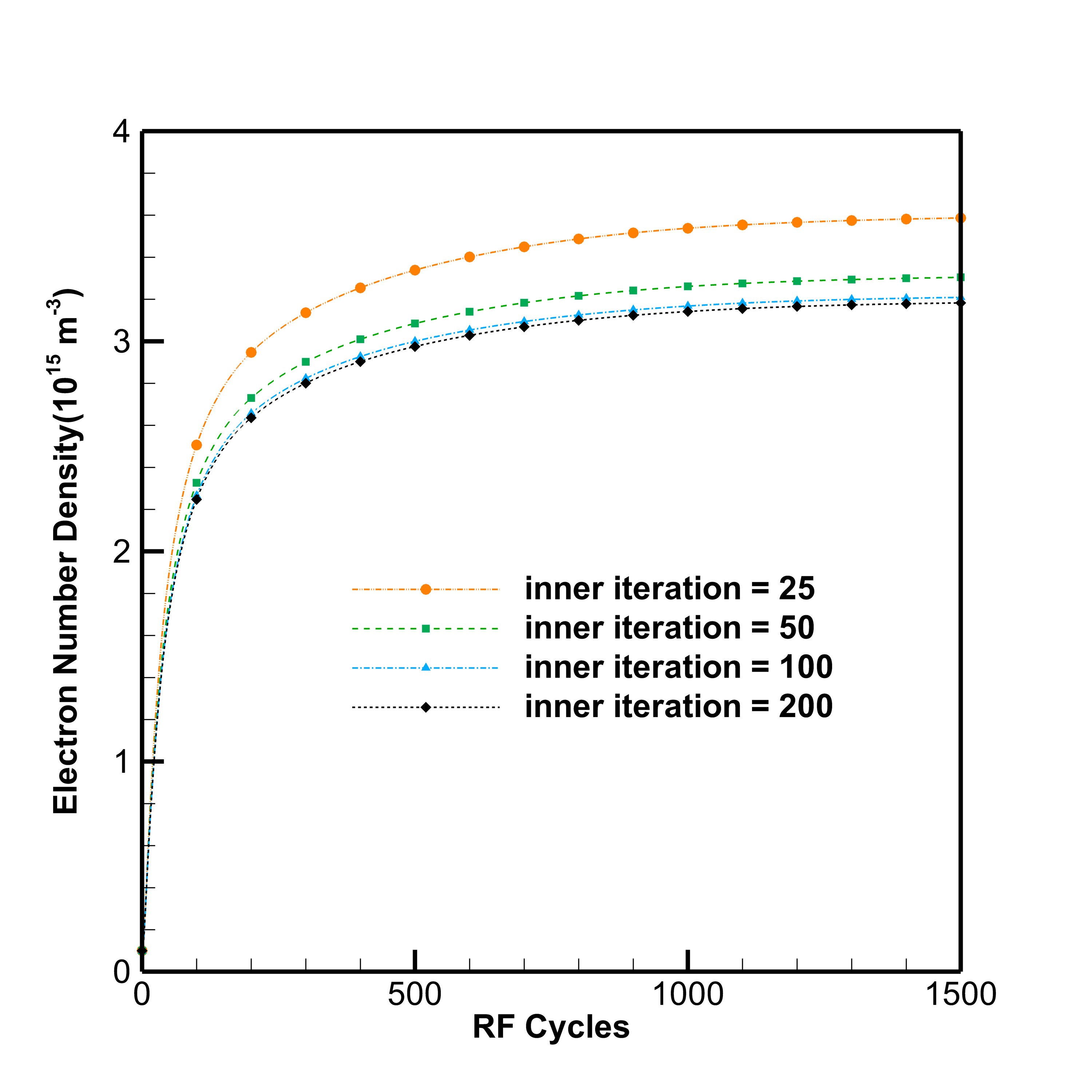}
    \end{minipage}\hfill
    \begin{minipage}{0.48\textwidth}
        \centering
        \includegraphics[width=\linewidth]{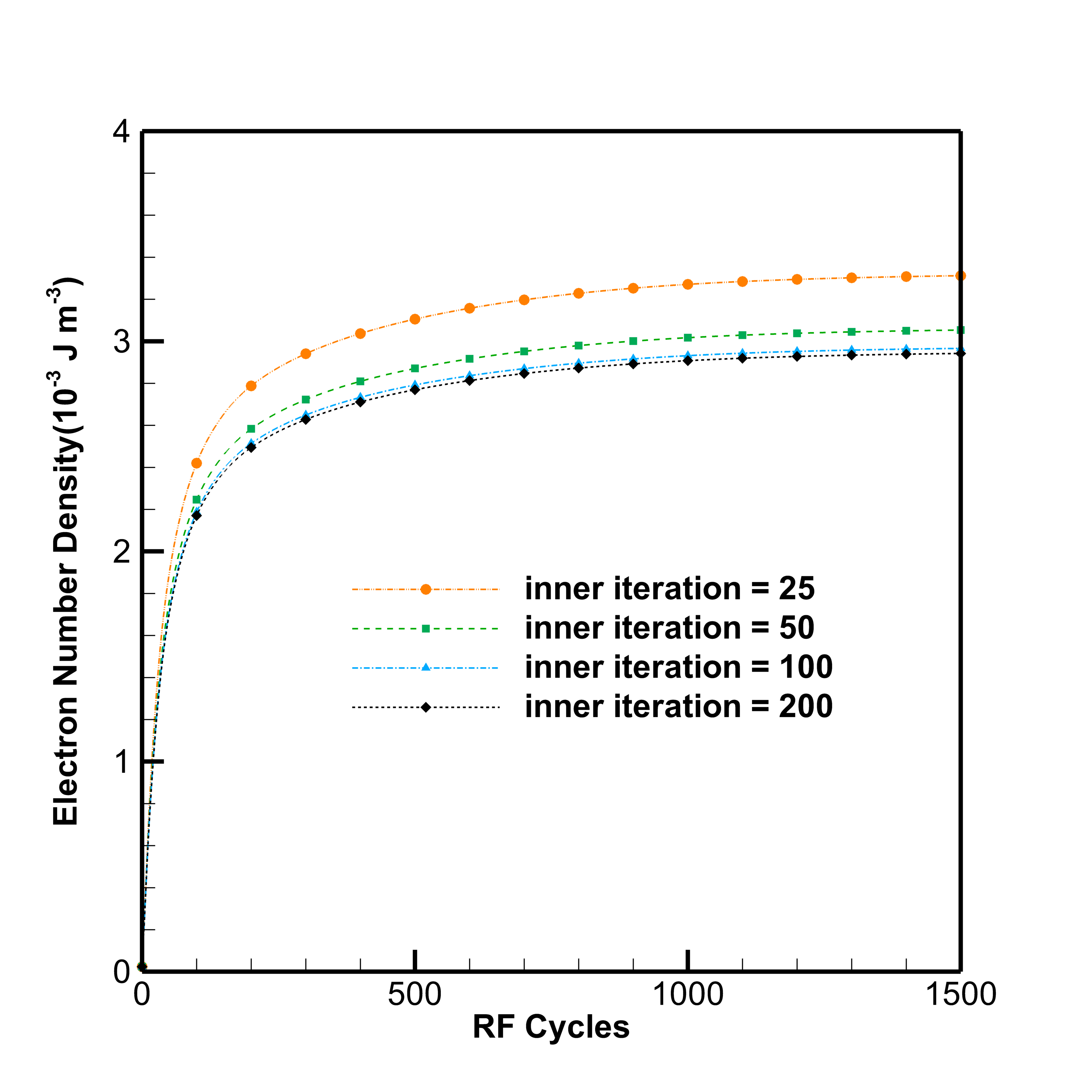}
    \end{minipage}
    \caption{Effect of the maximum number of pseudo-time inner iterations on the cycle-to-cycle evolution of the spatially averaged electron number density (left) and electron energy density (right), sampled at the beginning of each RF cycle.}
    \label{fig:inner_independence}
\end{figure}

Figure~\ref{fig:inner_independence} compares the  spatially averaged electron number density and electron energy density at the beginning of each RF cycle obtained  with different maximum numbers of pseudo-time inner iterations. The cases with 25 and 50 inner iterations still exhibit noticeable deviations, indicating insufficient nonlinear relaxation within each physical time step. In contrast, the solutions obtained with 100 and 200 inner iterations are already very close. At the 1500th RF cycle, the averaged electron number density differs by approximately 0.82\% between the 100- and 200-inner-iteration cases, while the corresponding difference in the averaged electron energy density is approximately 0.82\%. 

These observations indicate that approximately 200 inner iterations are required to nearly fully relax the BDF residual for the present benchmark, whereas 100 inner iterations already provide a sufficiently accurate periodic solution in terms of the monitored spatially averaged quantities. Using 200 inner iterations at every physical time step would considerably increase the computational cost, since each inner iteration involves the electrostatic potential update, flux and source-term evaluations, and block LU-SGS relaxation. In the present calculations, increasing the maximum number of inner iterations from 100 to 200 increases the wall-clock time at the 1500th RF cycle from 1404~s to 2679~s, corresponding to a cost increase of approximately 1.9 times. Therefore, a maximum of 100 pseudo-time inner iterations is adopted in the subsequent simulations as a practical compromise between inner-iteration convergence and computational efficiency.

\begin{figure}[htbp]
    \centering
    \begin{subfigure}[b]{0.48\textwidth}
        \centering
        \includegraphics[width=\textwidth]{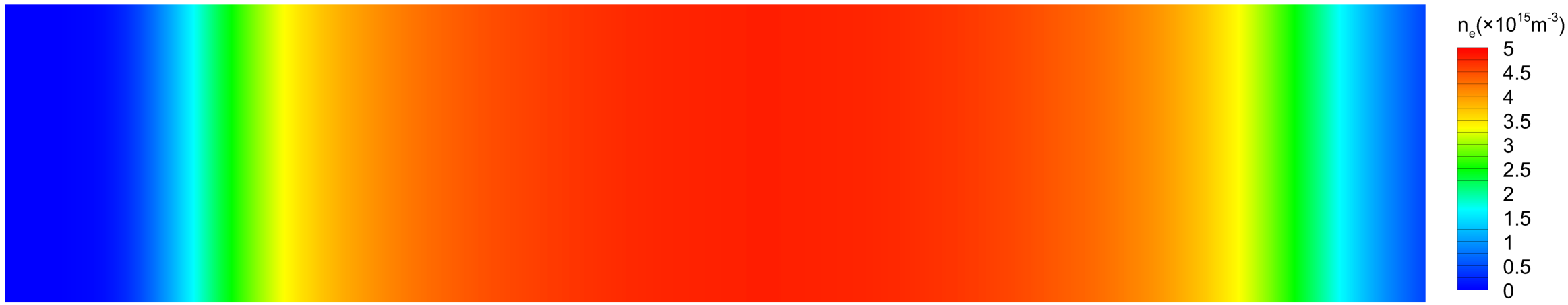}
        \caption{$t/T=0$}
    \end{subfigure}
    \hfill
    \begin{subfigure}[b]{0.48\textwidth}
        \centering
        \includegraphics[width=\textwidth]{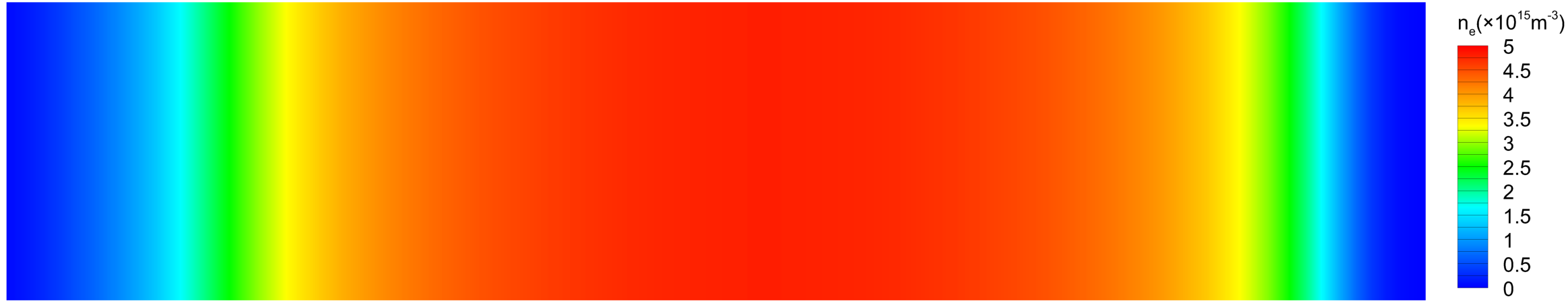}
        \caption{$t/T=0.25$}
    \end{subfigure}

    \vspace{0.5em}

    \begin{subfigure}[b]{0.48\textwidth}
        \centering
        \includegraphics[width=\textwidth]{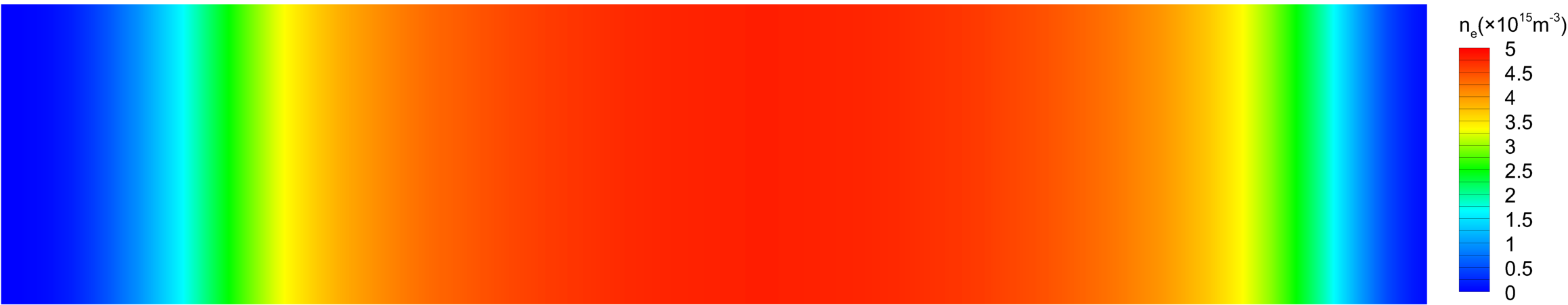}
        \caption{$t/T=0.50$}
    \end{subfigure}
    \hfill
    \begin{subfigure}[b]{0.48\textwidth}
        \centering
        \includegraphics[width=\textwidth]{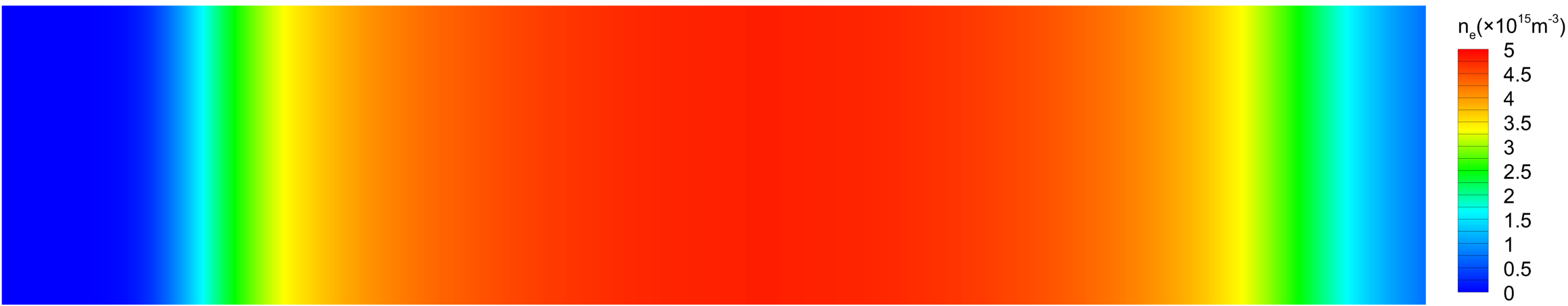}
        \caption{$t/T=0.75$}
    \end{subfigure}

    \caption{Phase-resolved contours of electron number density in the one-dimensional benchmark case represented on the two-dimensional slab domain, computed with pseudo-CFL $=10000$, $\Delta t = T/100$, and 100 inner iterations per physical time step.}
    \label{fig:ne_contours_1dcase}
\end{figure}

Figure~\ref{fig:ne_contours_1dcase} presents the electron number density contours at four representative phases within one RF period, namely $t/T=0$, $0.25$, $0.50$, and $0.75$, obtained with the specified numerical parameters of pseudo-CFL $=10000$, $\Delta t = T/100$, and 100 inner iterations per physical time step. 

As expected for the present two-dimensional slab configuration constructed to reproduce a one-dimensional benchmark discharge, the electron number density varies primarily along the interelectrode direction, while the variation in the transverse direction remains weak. At all phases, a high-density plasma bulk is formed in the central region, whereas the electron number density decreases rapidly near the electrode-adjacent sheath regions. The overall contour patterns at different phases remain similar, indicating that the dominant temporal modulation occurs mainly in the near-sheath regions, while the bulk plasma remains relatively stable over the RF cycle. 

To further assess the accuracy of the present solution, one-dimensional profiles are extracted along the centerline of the computational domain and compared with the digitized reference data of Lymberopoulos and Economou~\cite{lymberopoulos1993fluid}. The spatial coordinate is normalized by the electrode spacing ($\xi=x/L_x$).

\begin{figure}[htbp]
\centering

\begin{subfigure}[b]{0.48\textwidth}
    \centering
    \includegraphics[width=\textwidth]{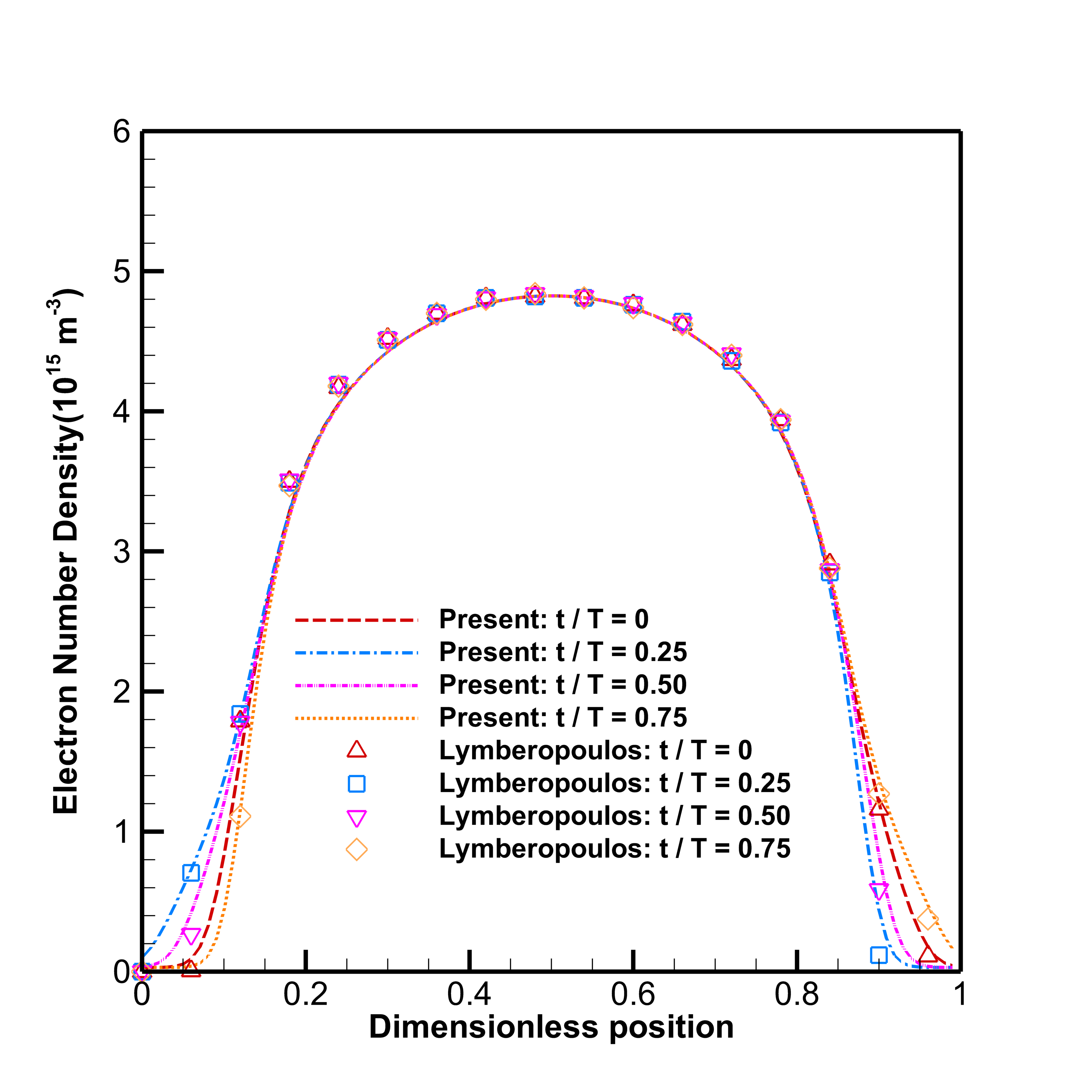}
    \caption{Electron number density profiles at four RF phases.}
    \label{fig:electron_profile_comparison}
\end{subfigure}
\begin{subfigure}[b]{0.48\textwidth}
    \centering
    \includegraphics[width=\textwidth]{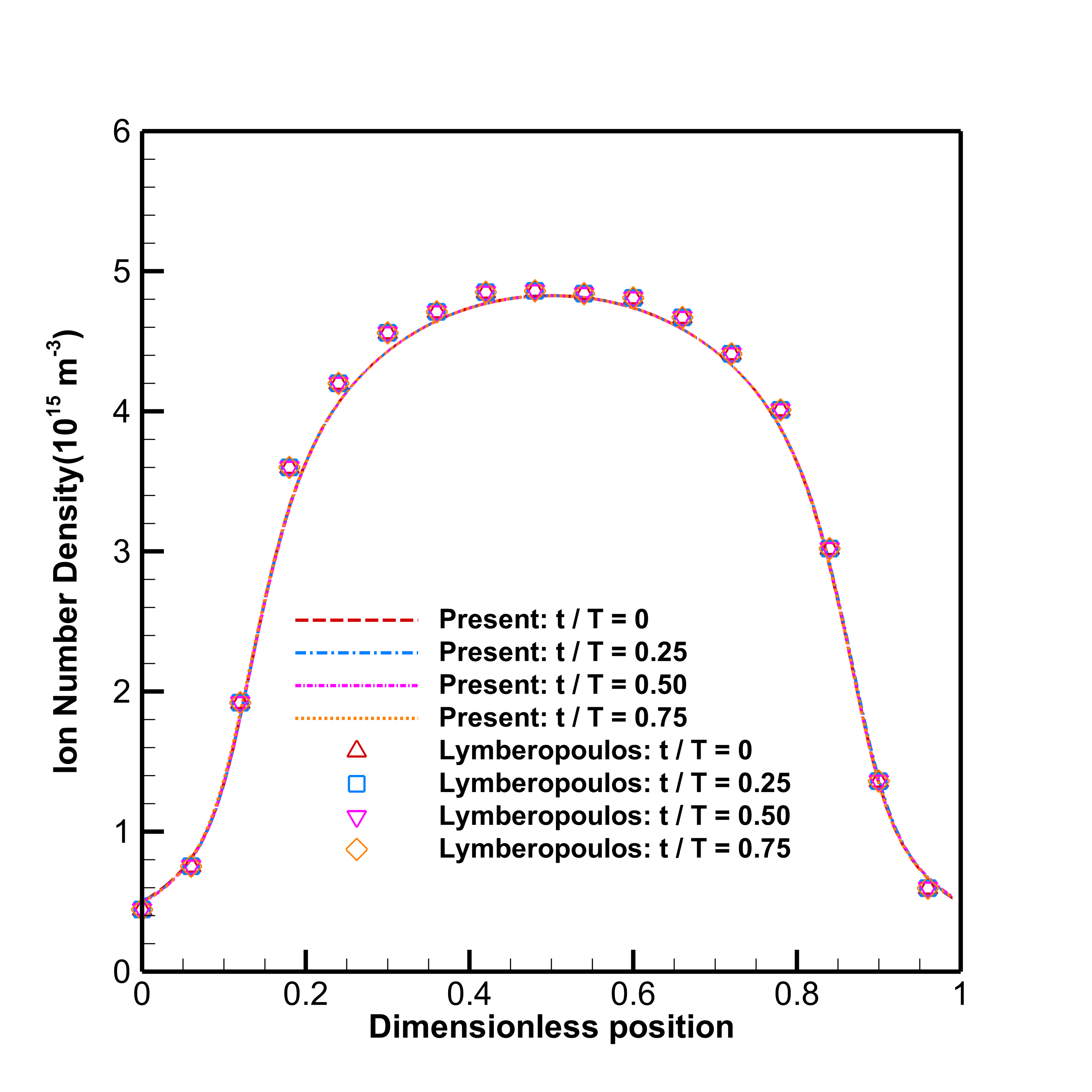}
    \caption{Ion number density profile.}
    \label{fig:ion_profile_comparison}
\end{subfigure}

\caption{Comparison of charged-particle density profiles with the digitized reference data of Lymberopoulos and Economou~\cite{lymberopoulos1993fluid}. Lines denote the present finite-volume results, while symbols represent the reference data.}
\label{fig:benchmark_profiles}

\end{figure}

Figure~\ref{fig:benchmark_profiles} compares the electron and ion number density profiles. For the electron number density (Fig.~\ref{fig:electron_profile_comparison}), four representative phases within one RF cycle, namely $t/T = 0$, $0.25$, $0.50$, and $0.75$, are shown. The present finite-volume results successfully reproduce the overall shape of the reference profiles, capturing both the high-density bulk region and the rapid density drop near the electrode-adjacent sheaths. The phase dependence of the electron  number  density is also accurately resolved, with the most noticeable discrepancies confined to the sheath edges where the spatial gradients are steepest. The corresponding ion number density profile is presented in Fig.~\ref{fig:ion_profile_comparison}. In contrast to the electrons, the ion number density exhibits a much weaker phase variation due to the slower ion response over an RF cycle. The present result closely follows the reference distribution across the entire interelectrode gap. These comparisons confirm that the proposed dual-time finite-volume solver accurately captures the primary charged-particle balance of the benchmark RF argon discharge.

A quantitative comparison of the spatially averaged charged-particle densities is provided in Table~\ref{tab}, which also includes the results obtained with a larger physical time step ($\Delta t = T/50$) for comparison. 

\begin{table}[h]
\centering
\caption{Comparison of spatially averaged charged-particle densities with the reference data.}
\label{tab}
\begin{tabular}{llcc}
\toprule
Quantity & Case & Average density ($10^{15}~\mathrm{m}^{-3}$) & Relative error \\
\midrule
$\langle n_e\rangle$
& Reference~\cite{lymberopoulos1993fluid}
& 3.2800
& -- \\
& Present ($\Delta t = T/100$)
& 3.2223
& $1.76\%$ \\
& Present ($\Delta t = T/50$)
& 3.2841
& $0.13\%$ \\
\midrule
$\langle n_i\rangle$
& Reference~\cite{lymberopoulos1993fluid}
& 3.4100
& -- \\
& Present ($\Delta t = T/100$)
& 3.3498
& $1.77\%$ \\
& Present ($\Delta t = T/50$)
& 3.4126
& $0.08\%$ \\
\bottomrule
\end{tabular}
\end{table}

Compared to the reference values reported by Lymberopoulos and Economou~\cite{lymberopoulos1993fluid}, the present solution obtained with $\Delta t = T/100$ yields relative errors of $1.76\%$ for the electron number density and $1.77\%$ for the ion number density. For comparison, a larger physical time step ($\Delta t = T/50$) fortuitously yields spatially averaged densities that are numerically closer to the reference, resulting in relative errors of $0.13\%$ and $0.08\%$, respectively.

These minor discrepancies between the present solution and the 1993 reference data naturally arise from fundamental differences in the numerical methodologies. Specifically, the reference calculation employed a one-dimensional Galerkin finite-element method coupled with a variable-time-step implicit solver, whereas the present approach relies on a cell-centered finite-volume discretization. Furthermore, the absence of reported details in the reference regarding the transient time-step sizes and the precise interpolation of tabulated electron-impact rate coefficients further accounts for these small quantitative deviations. Nevertheless, the excellent overall agreement in density magnitudes, profile shapes, and spatially averaged values confirms that the present dual-time finite-volume solver reliably captures the essential physics of the benchmark RF argon discharge.

\begin{figure}[htbp]
    \centering
    \includegraphics[width=0.7\textwidth]{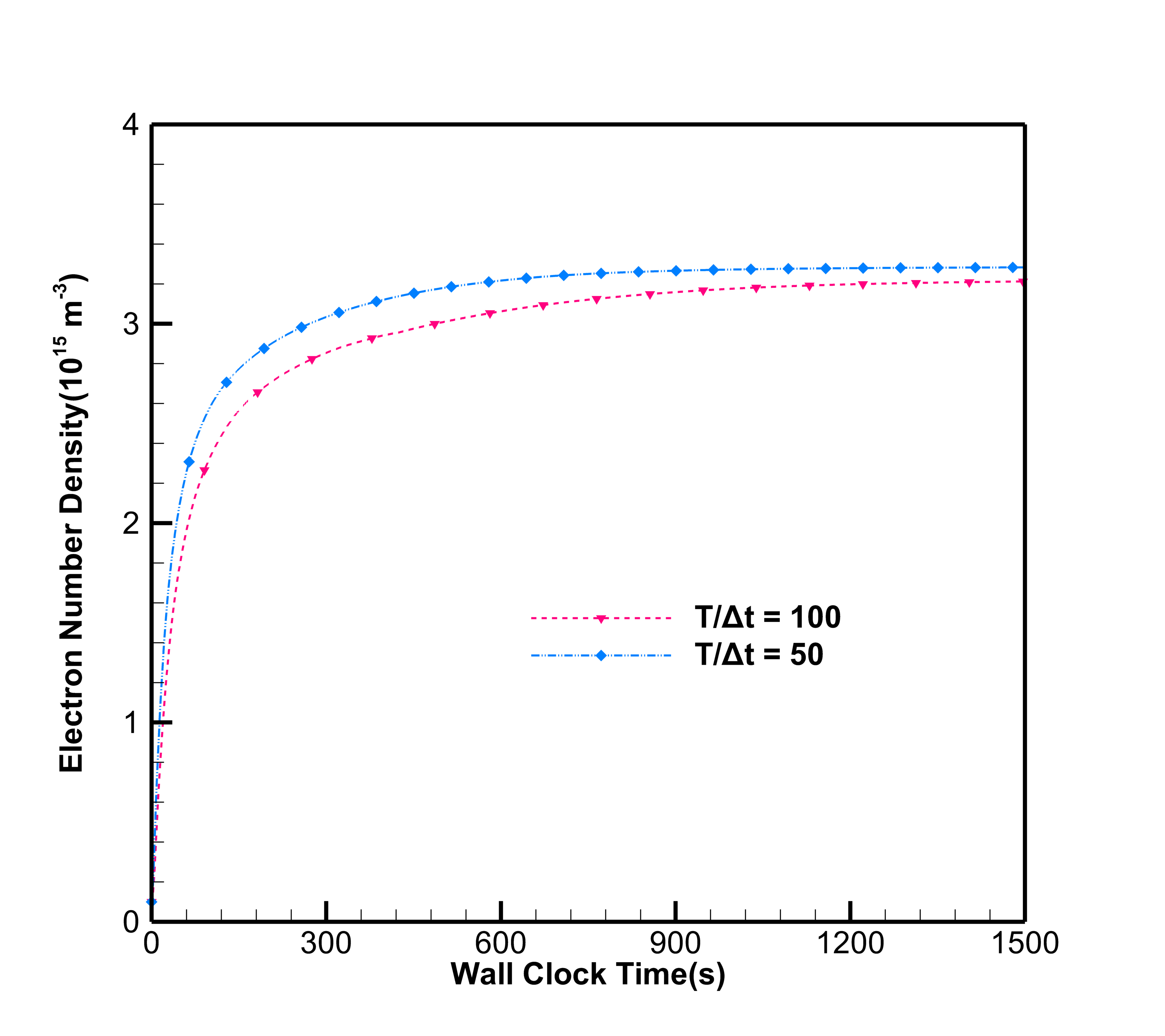} 
    \caption{Convergence history of the spatially averaged electron number density as a function of wall-clock time for different physical time steps. The simulation is performed using a single-threaded execution.}
    \label{fig:wall_clock_time}
\end{figure}

Finally, the computational efficiency of the proposed dual-time finite-volume solver is assessed. Figure~\ref{fig:wall_clock_time} illustrates the convergence history of the spatially averaged electron number density as a function of wall-clock time. To provide a clear baseline for the computational cost, the simulation was performed using a single-threaded serial execution on an Intel Core Ultra 9 285H processor. 

As shown, the macroscopic plasma dynamics rapidly evolve and reach a fully converged periodic steady state within approximately $1000~\mathrm{s}$ of wall-clock time for the primary case ($\Delta t = T/100$). This fast turnaround time demonstrates the practical advantage of the present implicit framework. By solving the implicit BDF residual through pseudo-time iterations, the solver successfully avoids the restrictive stability limits of conventional explicit schemes, thereby delivering an optimal balance between temporal accuracy, numerical robustness, and overall computational efficiency.

\subsection{Two-dimensional RF CCP simulations}
\subsubsection{Numerical Setup}

\begin{figure}[htbp]
    \centering
    \includegraphics[width=0.85\textwidth]{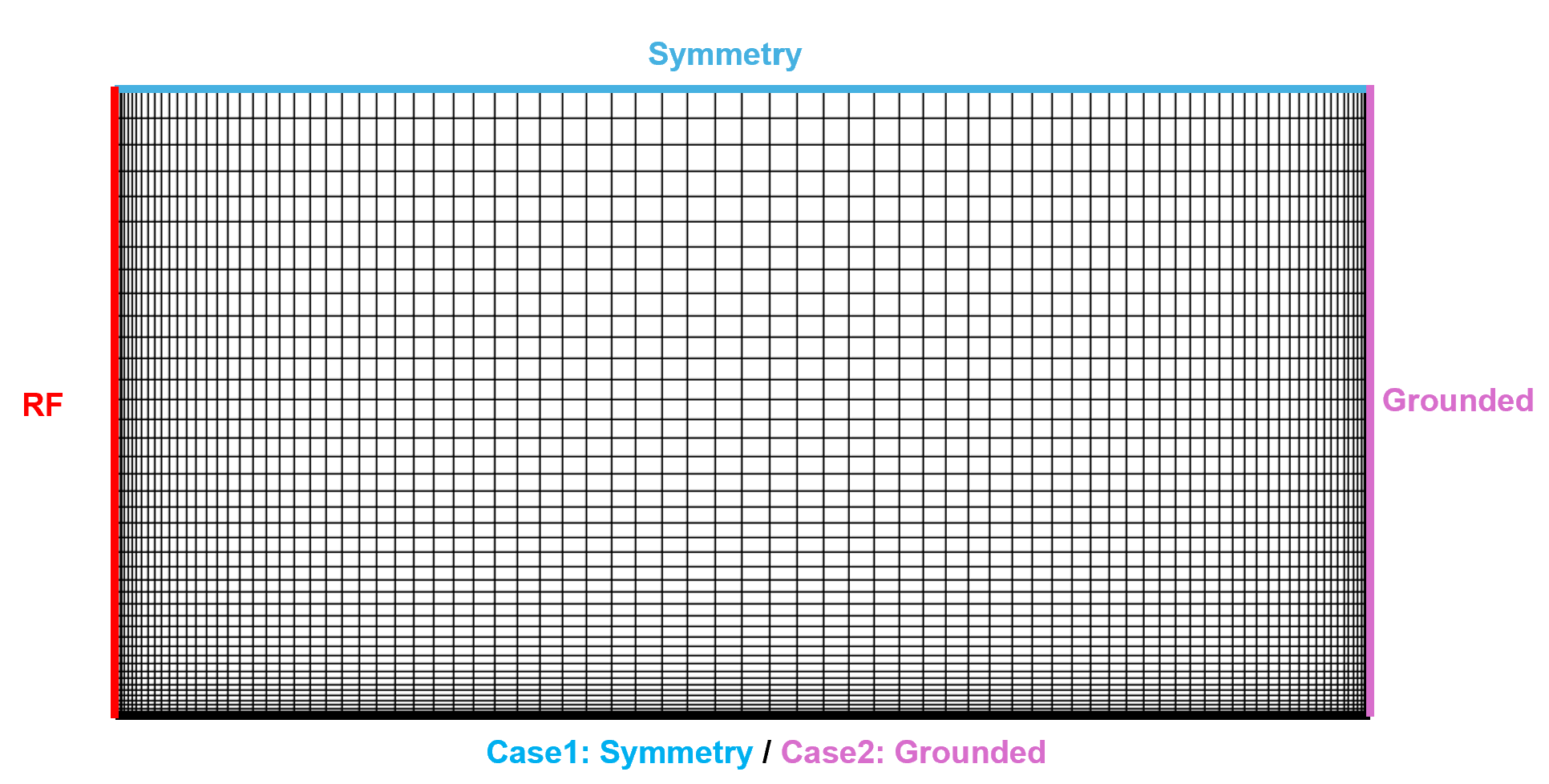}
    \caption{Schematic diagram of the computational domain and boundary conditions for the two-dimensional RF CCP simulations. Case 1 features symmetry conditions on both top and bottom boundaries, while Case 2 employs a grounded bottom boundary.}
    \label{fig:2d_schematic}
\end{figure}

To further demonstrate the multidimensional capability and robustness of the present dual-time finite-volume solver, the computations are extended to a two-dimensional spatial domain. The computational domain is a rectangle with a length of $L_x = 0.0254~\mathrm{m}$ in the $x$-direction and a height of $L_y = L_x / 2 = 0.0127~\mathrm{m}$ in the $y$-direction. As schematically illustrated in Fig.~\ref{fig:2d_schematic}, the left boundary ($x=0$) is powered by the identical RF voltage ($V_0 = 100~\mathrm{V}$, phase $0$) used in the 1D benchmark, while the right boundary ($x=L_x$) is strictly grounded.

To systematically investigate multidimensional effects, two specific configurations are considered:
\begin{itemize}
	\item \textbf{Case 1:} Zero-gradient symmetry conditions are imposed on both the top ($y=L_y$) and bottom ($y=0$) boundaries. This effectively eliminates transverse gradients, reducing the setup to a quasi-one-dimensional discharge. Notably, even though the physical solution degenerates to one dimension, the underlying mesh in the $y$-direction remains non-uniform, thereby serving as a robust baseline for verifying the solver's multi-dimensional implementation on non-uniform grids.
    \item \textbf{Case 2:} The top boundary remains a symmetry plane, but the bottom boundary is grounded ($\phi = 0$). The inclusion of a transverse grounded wall introduces spatial asymmetries and drives genuine two-dimensional charged-particle transport and sheath dynamics.
\end{itemize}

The spatial discretization employs a $90 \times 45$ non-uniform Cartesian cell mesh. In the interelectrode ($x$) direction, the node distribution is identical to that of the 1D benchmark, ensuring adequate resolution of the steep gradients within the electrode sheaths. In the transverse ($y$) direction, the node distribution is generated using the identical formulation as the $x$-direction based on the full length $L_x$, but only the lower half of the grid is retained to form the $L_y$ domain. This truncation naturally preserves the mesh clustering near the bottom boundary ($y=0$), ensuring sufficient resolution for the near-wall sheath structures in Case 2.

Consistent with the physical time-step verification, the temporal resolution is maintained at $\Delta t = T/100$, and a maximum of 100 pseudo-time inner iterations is performed per physical time step. Based on the preceding inner-iteration convergence assessment, a pseudo-CFL number of $3000$ is adopted. This corresponds to a pseudo-time step of approximately $1.7 \times 10^{-11}~\mathrm{s}$, which is found to provide stable nonlinear relaxation and satisfy the convergence requirements for the 2D computations without incurring unnecessary computational overhead.

\subsubsection{Numerical Results}

\begin{figure}[h]
\centering
\includegraphics[width=0.7\textwidth]{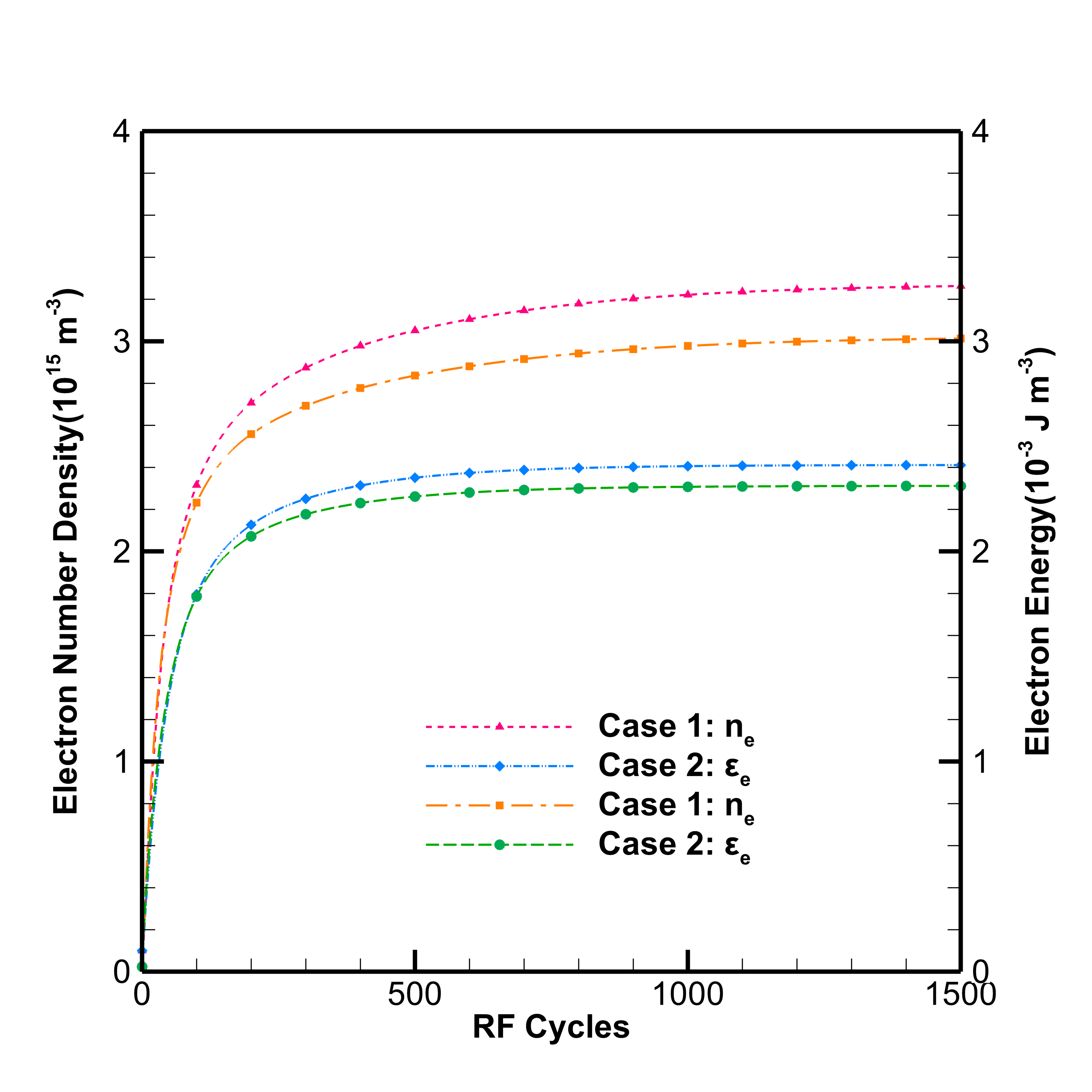}
\caption{Temporal evolution of spatially averaged quantities over RF cycles for the two-dimensional calculations.}
\label{fig:case2d_convergence}
\end{figure}

We first examine the evolution of the macroscopic spatial averages. Figure~\ref{fig:case2d_convergence} illustrates the cycle-to-cycle temporal evolution of the spatially averaged quantities for both cases.

In both cases, the spatially averaged electron number density and electron energy density initially experience a  growth phase before gradually leveling off. A well-defined periodic state is successfully established after approximately 1000 RF cycles. Subsequently, a detailed quantitative comparison of the converged spatially averaged electron densities is provided in Table~\ref{tab:case2d_average_density}.

\begin{table}[htbp]
\centering
\caption{Comparison of spatially averaged electron number density for the two-dimensional strip calculations.}
\label{tab:case2d_average_density}
\begin{tabular}{lcc}
\toprule
Case & Average density ($10^{15}~\mathrm{m}^{-3}$) & Relative difference \\
\midrule
Reference 1D solution~\cite{lymberopoulos1993fluid} & 3.280 & -- \\
Case 1: Transverse symmetry & 3.263 & $0.52\%$ \\
Case 2: Grounded lower wall & 2.411 & $26.49\%$ \\
\bottomrule
\end{tabular}
\end{table}

\begin{figure}[h]
\centering

\begin{subfigure}[b]{0.48\textwidth}
    \centering
    \includegraphics[width=\textwidth]{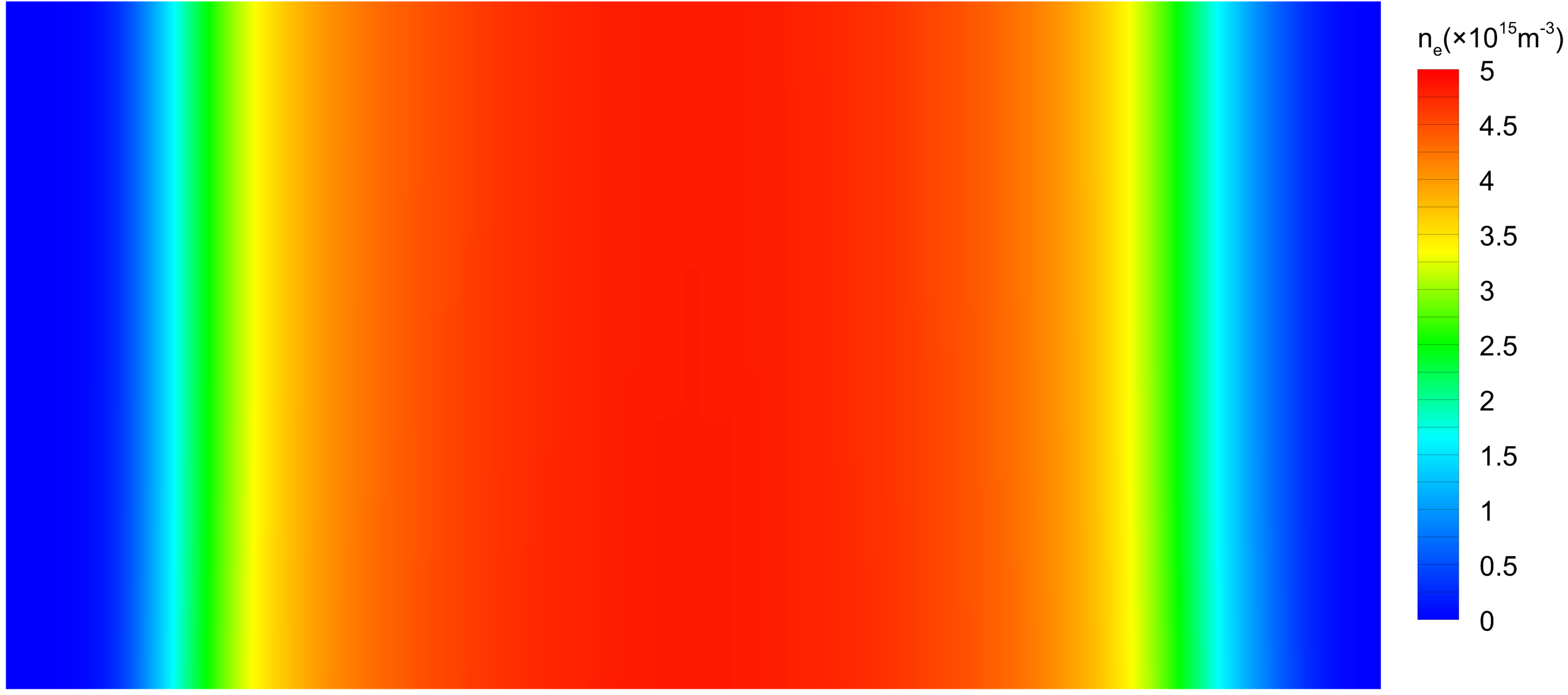}
    \caption{Electron number density $n_e$}
    \label{fig:case1_ne_t0}
\end{subfigure}
\hfill
\begin{subfigure}[b]{0.48\textwidth}
    \centering
    \includegraphics[width=\textwidth]{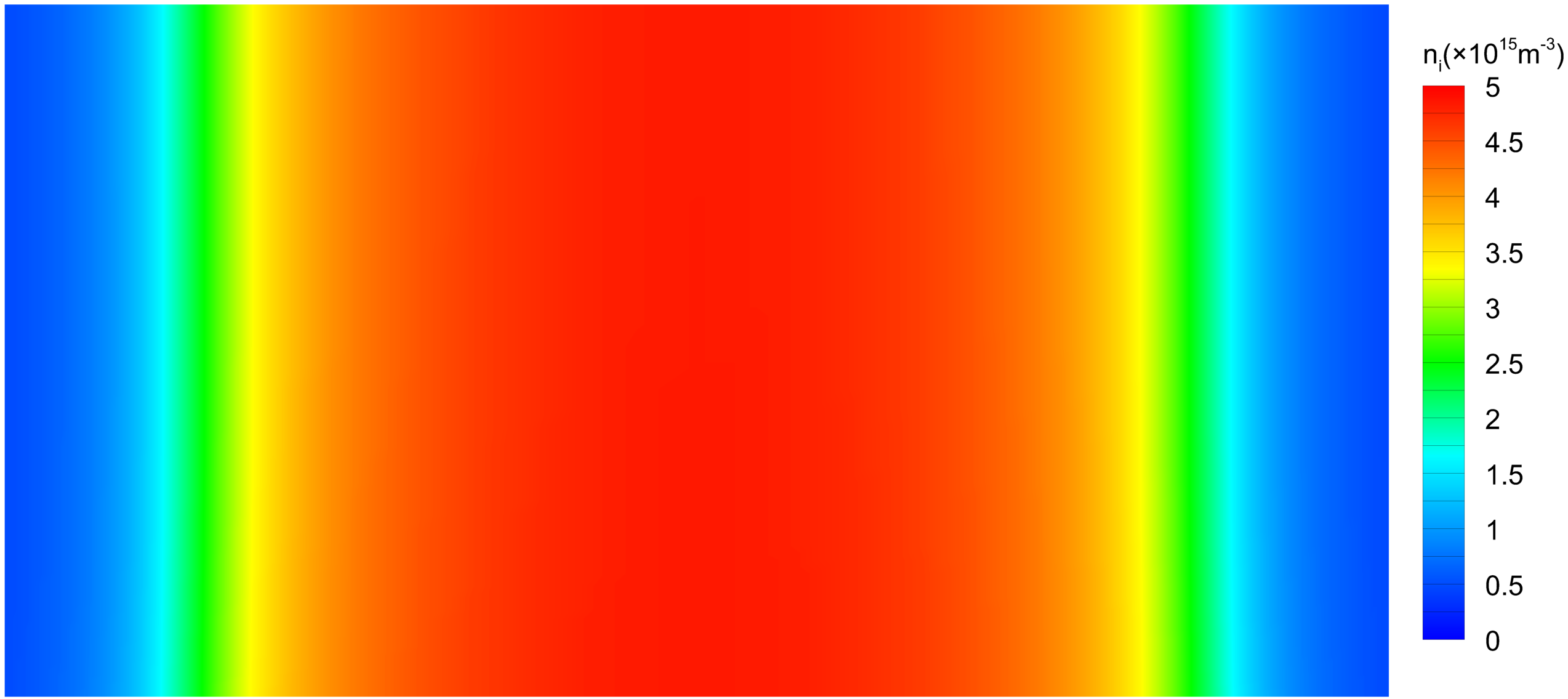}
    \caption{Ion number density $n_i$}
    \label{fig:case1_ni_t0}
\end{subfigure}

\vspace{1.5em}

\begin{subfigure}[b]{0.48\textwidth}
    \centering
    \includegraphics[width=\textwidth]{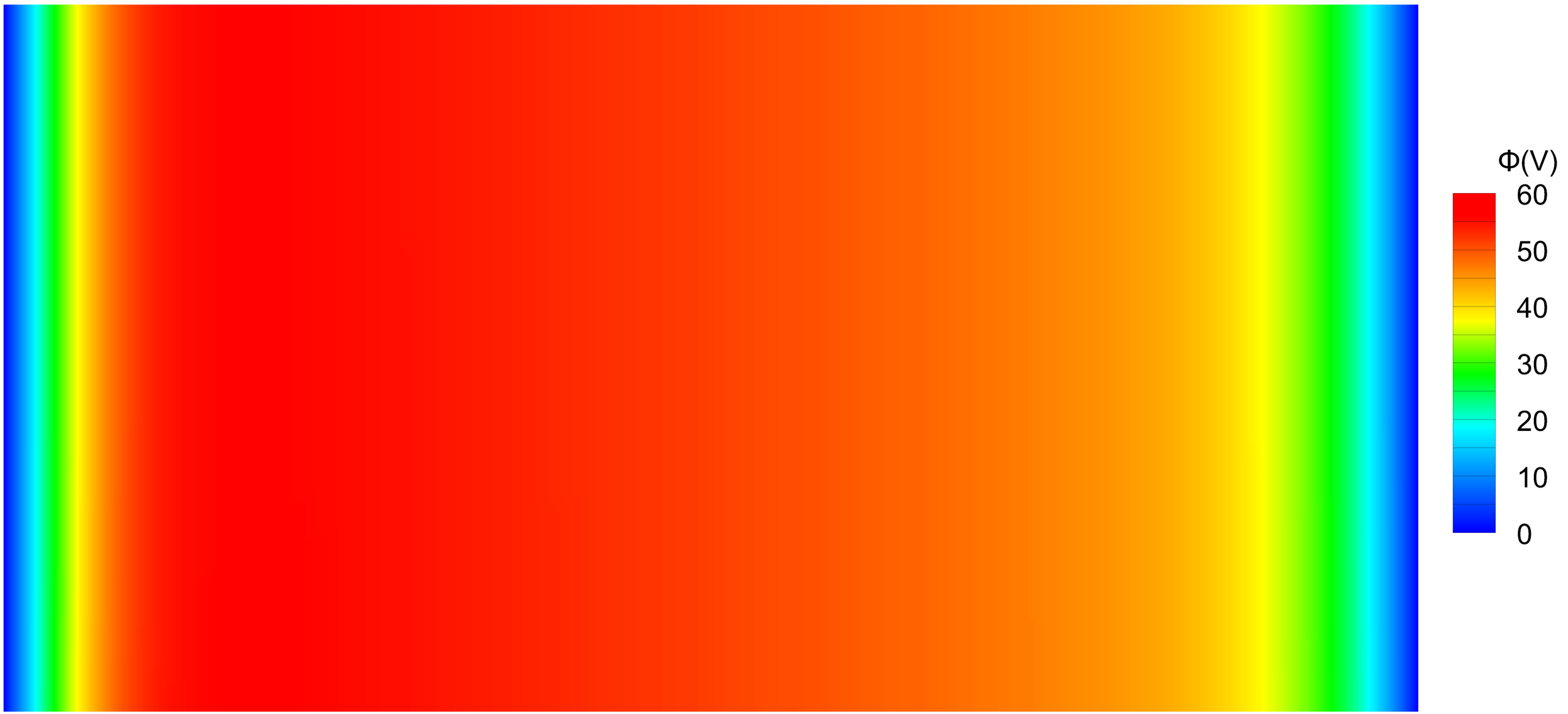}
    \caption{Electrostatic potential $\phi$}
    \label{fig:case1_phi_t0}
\end{subfigure}
\hfill
\begin{subfigure}[b]{0.48\textwidth}
    \centering
    \includegraphics[width=\textwidth]{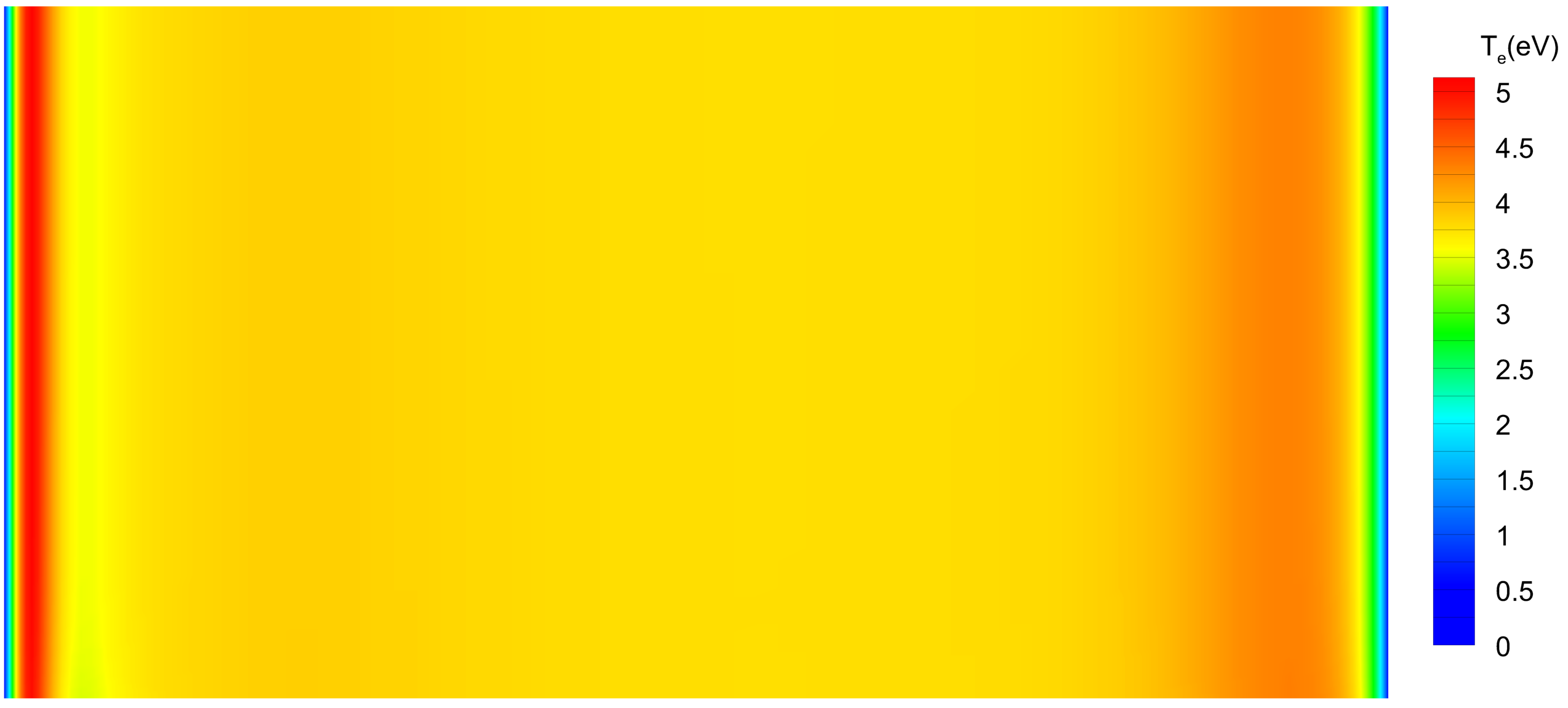}
    \caption{Electron temperature $T_e$}
    \label{fig:case1_Te_t0}
\end{subfigure}

\caption{Distributions of the main plasma variables in Case 1 at $t/T=0$..}
\label{fig:case1_contours_t0}

\end{figure}

For Case 1, the calculated spatially averaged electron number density yields a marginal relative difference of $0.52\%$ compared to the 1D reference value, confirming that the 2D solver accurately preserves the quasi-one-dimensional discharge physics. To provide a more intuitive visualization of the macroscopic field variables, Figure~\ref{fig:case1_contours_t0} presents the spatial distributions of the electron number density, ion number density, electrostatic potential, and electron temperature for Case 1 at the beginning of the final RF cycle ($t/T=0$).

As anticipated for a quasi-one-dimensional setup, all contour lines are strictly parallel to the $y$-axis, demonstrating the absence of transverse variations. The plasma bulk exhibits the highest charged-particle number densities, which rapidly decay near both electrodes due to the formation of longitudinal sheaths. Similarly, the electrostatic potential and electron temperature distributions are entirely dominated by interelectrode gradients, with strong spatial modulations tightly confined to the vicinity of the RF and grounded electrodes. These contour plots visually corroborate that the multi-dimensional calculation recovers the underlying one-dimensional physical structure, providing a rigorous baseline for analyzing the subsequent case.

In contrast, the spatially averaged electron number density in Case 2 drops significantly to $2.411\times 10^{15}~\mathrm{m}^{-3}$. This $26.49\%$ reduction is physically expected: the grounded lower wall introduces a substantial transverse charged-particle sink, fundamentally altering the global balance between ionization production and wall loss. 

\begin{figure}[h]
\centering
\begin{subfigure}[b]{0.48\textwidth}
    \centering
    \includegraphics[width=\textwidth]{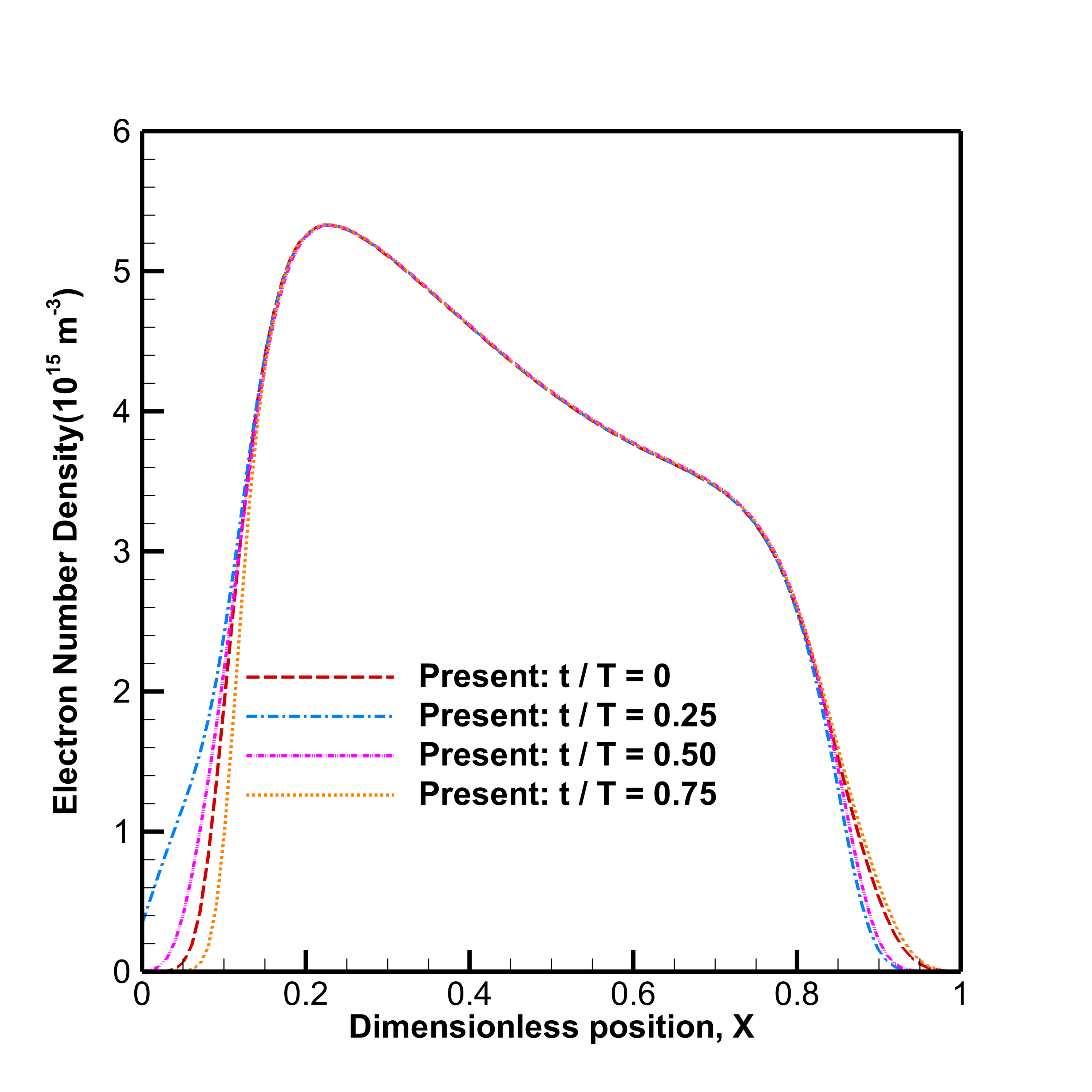}
    \caption{Horizontal profile at $y=L_y$.}
    \label{fig:case2_ne_x}
\end{subfigure}
\hfill
\begin{subfigure}[b]{0.48\textwidth}
    \centering
    \includegraphics[width=\textwidth]{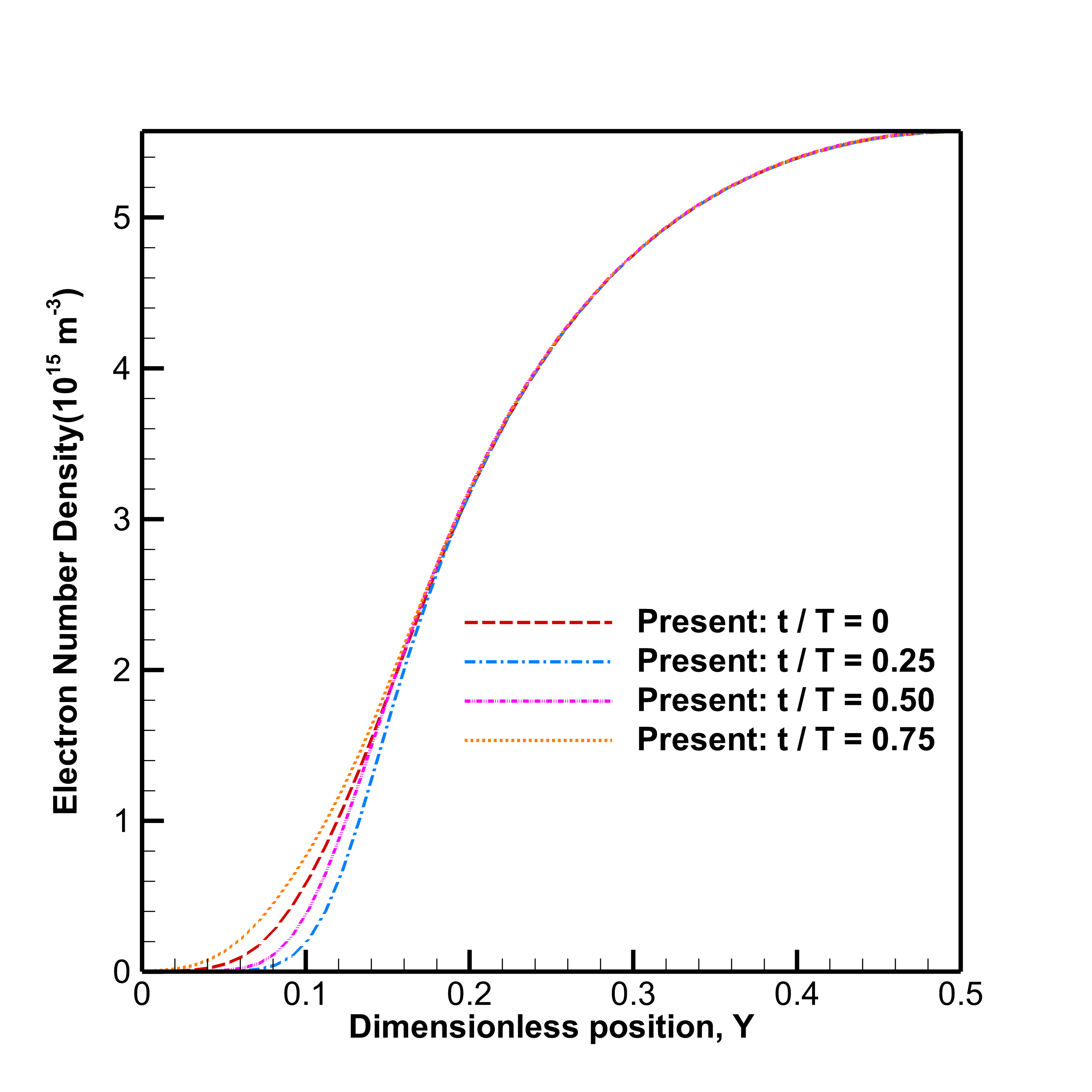}
    \caption{Vertical profile at $x=L_x/2$.}
    \label{fig:case2_ne_y}
\end{subfigure}
\caption{Phase-resolved one-dimensional electron number density profiles for Case 2.}
\label{fig:case2_1d_profiles}
\end{figure}

Subsequently, one-dimensional spatial profiles of the electron  number density are extracted and plotted in Fig.~\ref{fig:case2_1d_profiles} to demonstrate the physical mechanism behind this density reduction. The horizontal profile (Fig.~\ref{fig:case2_ne_x}) is taken along the upper symmetry boundary ($y=L_y$), while the vertical profile (Fig.~\ref{fig:case2_ne_y}) is extracted along the geometric centerline of the interelectrode gap ($x=L_x/2$). 

The horizontal profile (Fig.~\ref{fig:case2_ne_x}) still exhibits sharp density declines near the left and right electrodes; however, the high-density bulk region is no longer longitudinally symmetric. The addition of the grounded lower wall breaks the macroscopic electrical symmetry of the discharge, causing the electron number density distribution to skew noticeably toward the powered electrode, with its local maximum shifting to a dimensionless horizontal position of approximately $\xi = x/L_x = 0.2$. Furthermore, the overall magnitude of the density is visibly reduced compared to the quasi-one-dimensional case due to the continuous transverse particle loss. This transverse loss is explicitly confirmed by the vertical profile in Fig.~\ref{fig:case2_ne_y}. The electron number density peaks at the top symmetry plane but drops rapidly to near zero at the bottom boundary ($y=0$), indicating the formation of a strong transverse sheath. The continuous diffusion and subsequent recombination of electrons at this grounded lower wall represent the primary cause of both the diminished global average density and the broken longitudinal symmetry.

\begin{figure}[!h]
    \centering
    \includegraphics[width=0.7\textwidth]{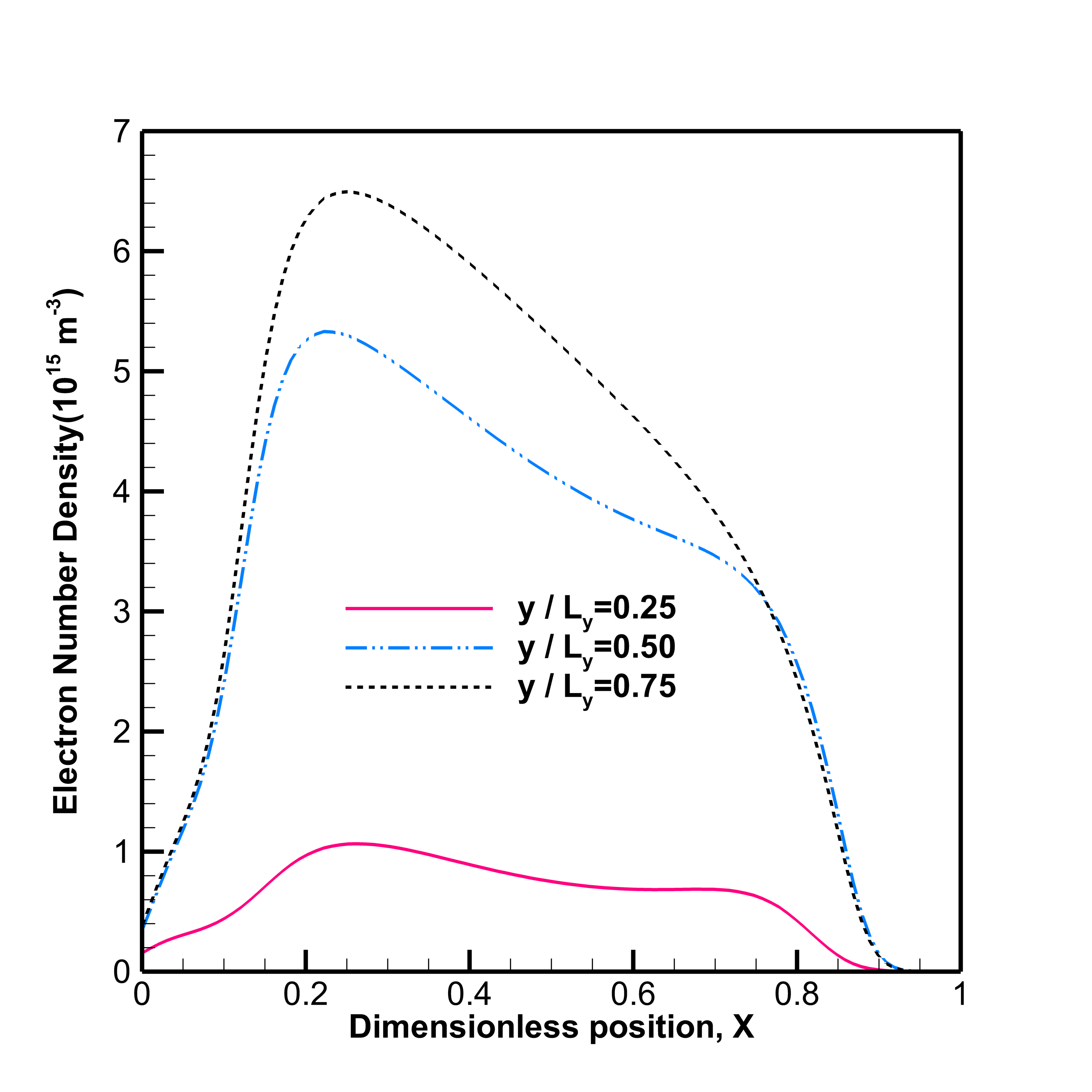}
    \caption{Horizontal electron number density profiles at different transverse locations in Case 2. }
    \label{fig:case2_ne_x_different_y}
\end{figure}

To further elucidate the spatial suppression caused by the grounded wall, Figure~\ref{fig:case2_ne_x_different_y} compares the horizontal electron number density profiles at three distinct transverse heights ($\zeta=y/L_y = 0.25, 0.50$, and $0.75$). 

As the observation plane moves closer to the grounded lower boundary (from $\zeta = 0.75$ down to $0.25$), the global magnitude of the electron number density is drastically diminished. While the macroscopic asymmetry---characterized by the density peak skewed toward the powered electrode ($\xi \approx 0.2$)---remains persistent across all heights, the bulk plasma region becomes increasingly depleted. This quantitative comparison demonstrates that the transverse particle sink does not merely affect the immediate vicinity of the bottom wall. Instead, it exerts a profound global influence, continuously draining charged particles and suppressing the plasma bulk across the entire interelectrode gap.

\begin{figure}[!h]
\centering
\begin{subfigure}[b]{0.48\textwidth}
    \centering
    \includegraphics[width=\textwidth]{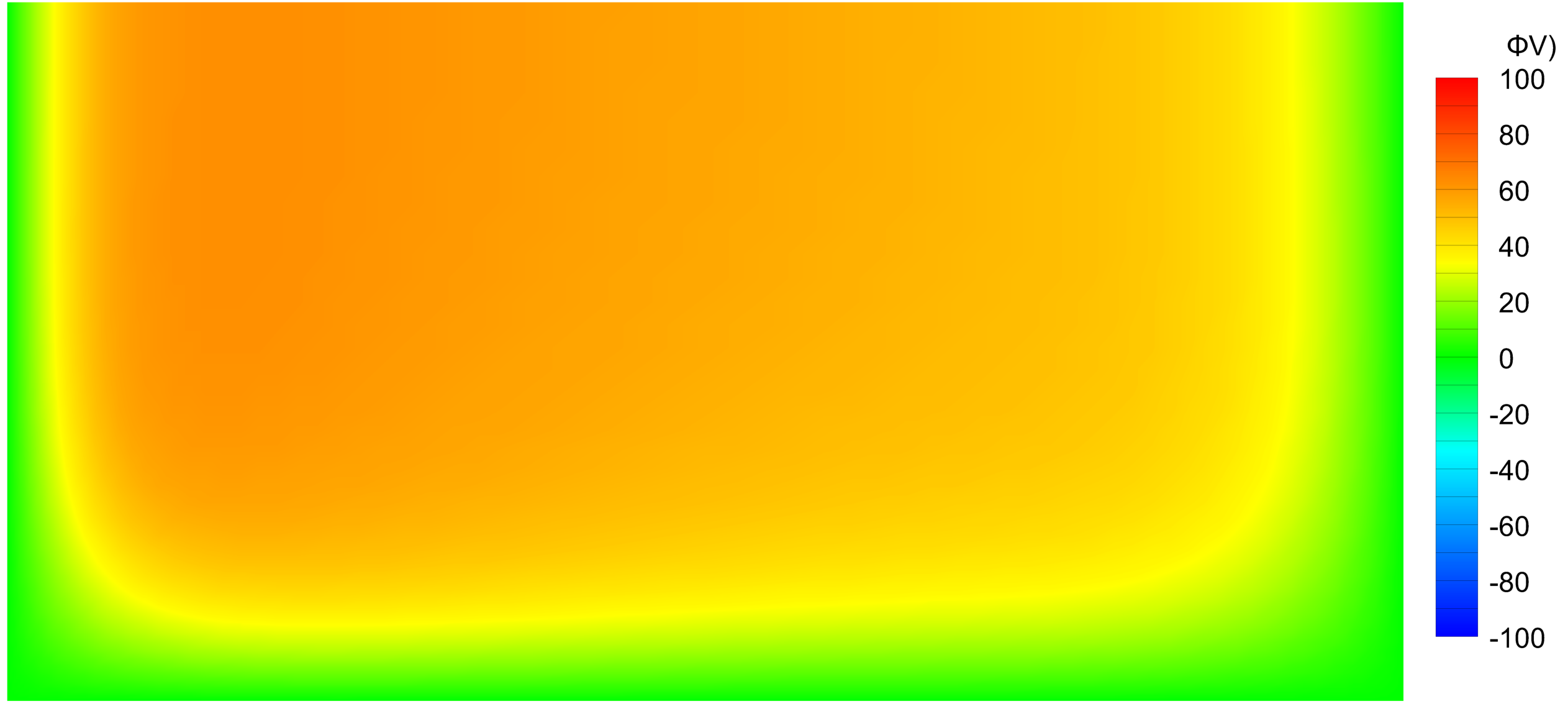}
    \caption{$t/T=0$}
\end{subfigure}
\hfill
\begin{subfigure}[b]{0.48\textwidth}
    \centering
    \includegraphics[width=\textwidth]{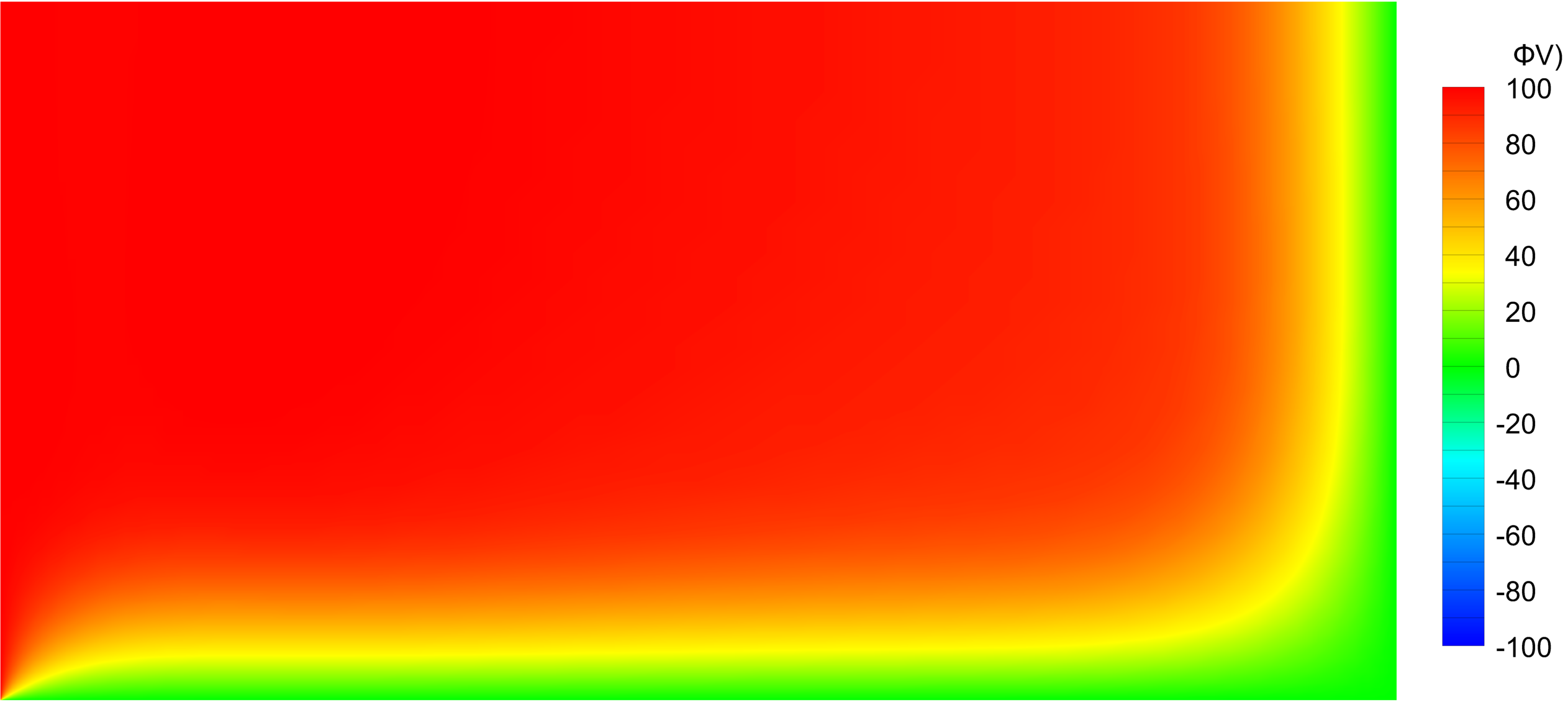}
    \caption{$t/T=0.25$}
\end{subfigure}

\vspace{1.0em}

\begin{subfigure}[b]{0.48\textwidth}
    \centering
    \includegraphics[width=\textwidth]{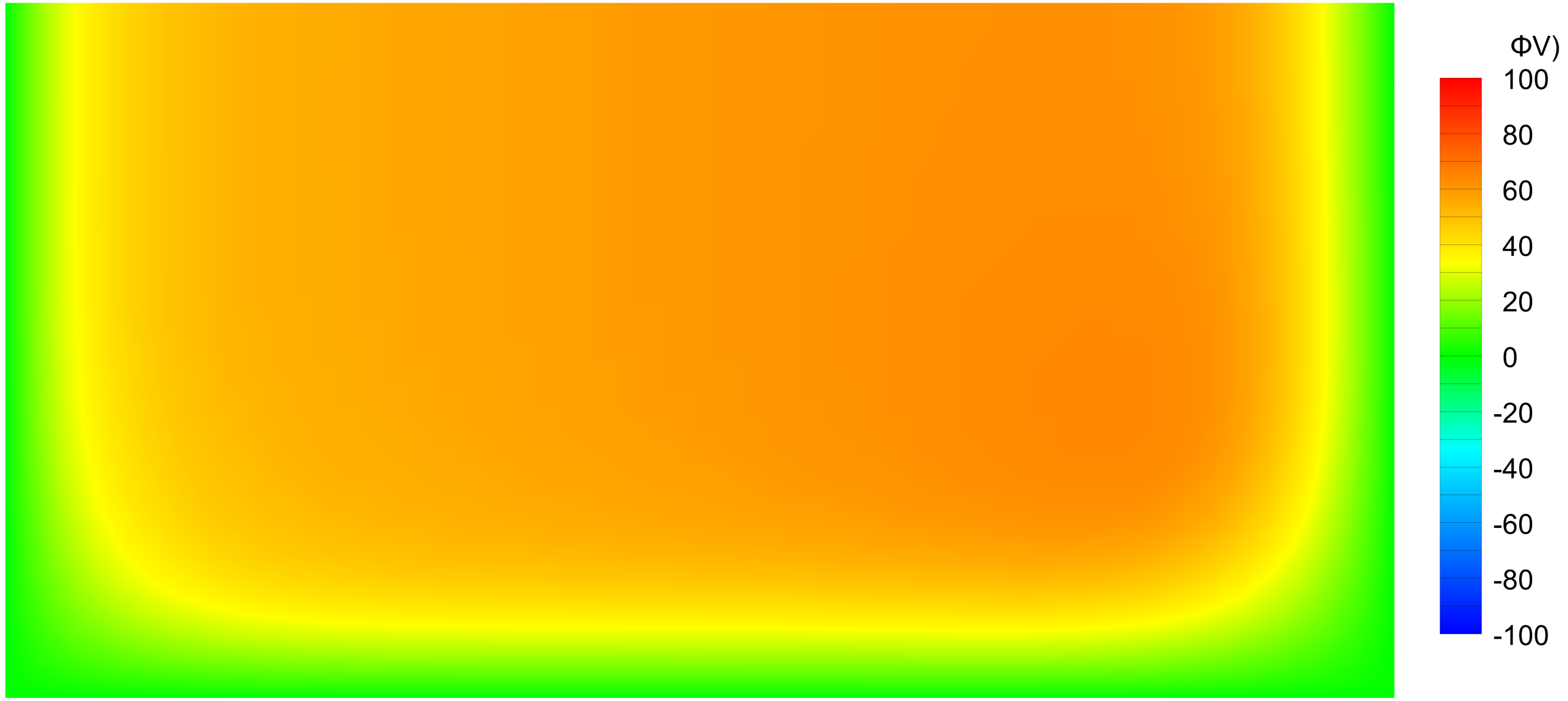}
    \caption{$t/T=0.50$}
\end{subfigure}
\hfill
\begin{subfigure}[b]{0.48\textwidth}
    \centering
    \includegraphics[width=\textwidth]{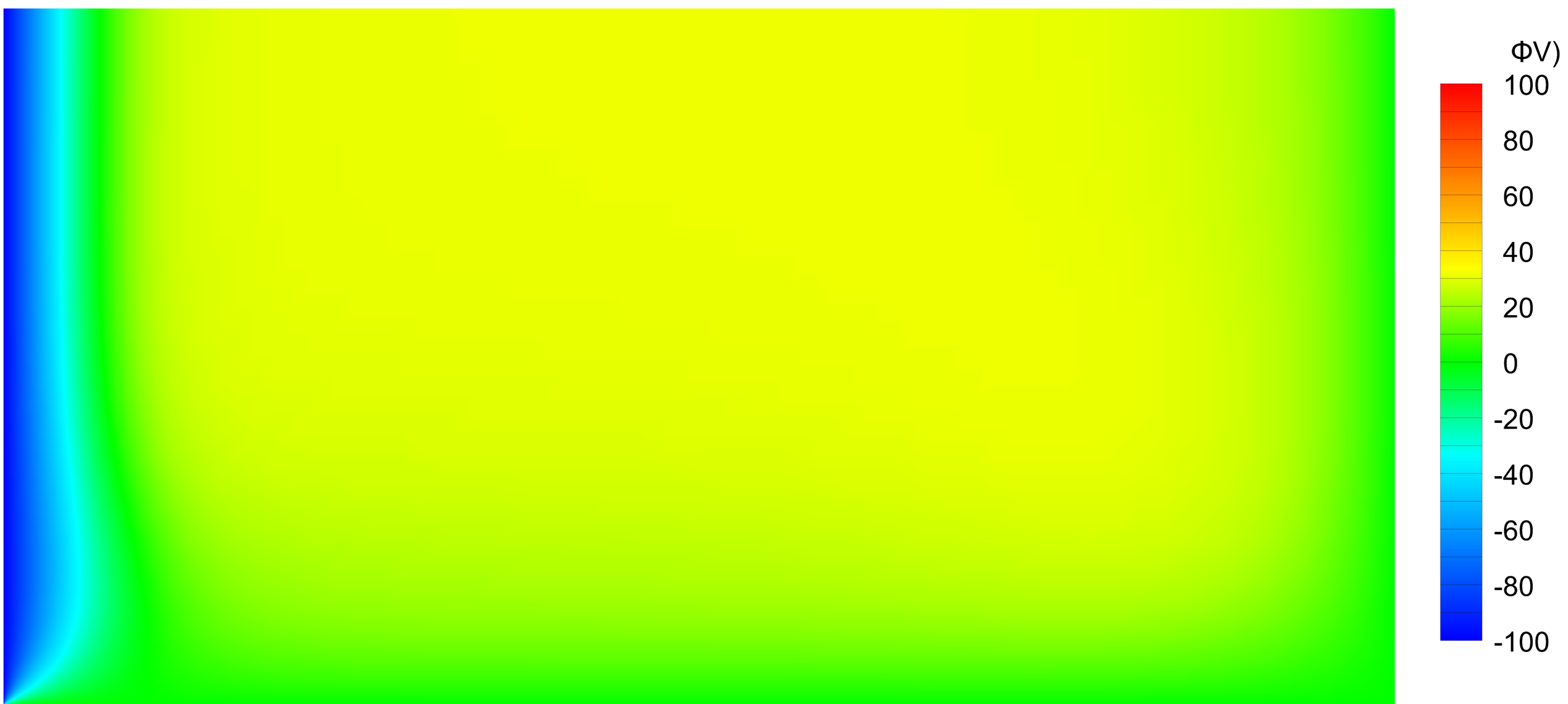}
    \caption{$t/T=0.75$}
\end{subfigure}
\caption{Phase-resolved contours of the electrostatic potential $\phi$ in Case 2.}
\label{fig:case2_phi_contours}
\end{figure}

\begin{figure}[!h]
\centering
\begin{subfigure}[b]{0.48\textwidth}
    \centering
    \includegraphics[width=\textwidth]{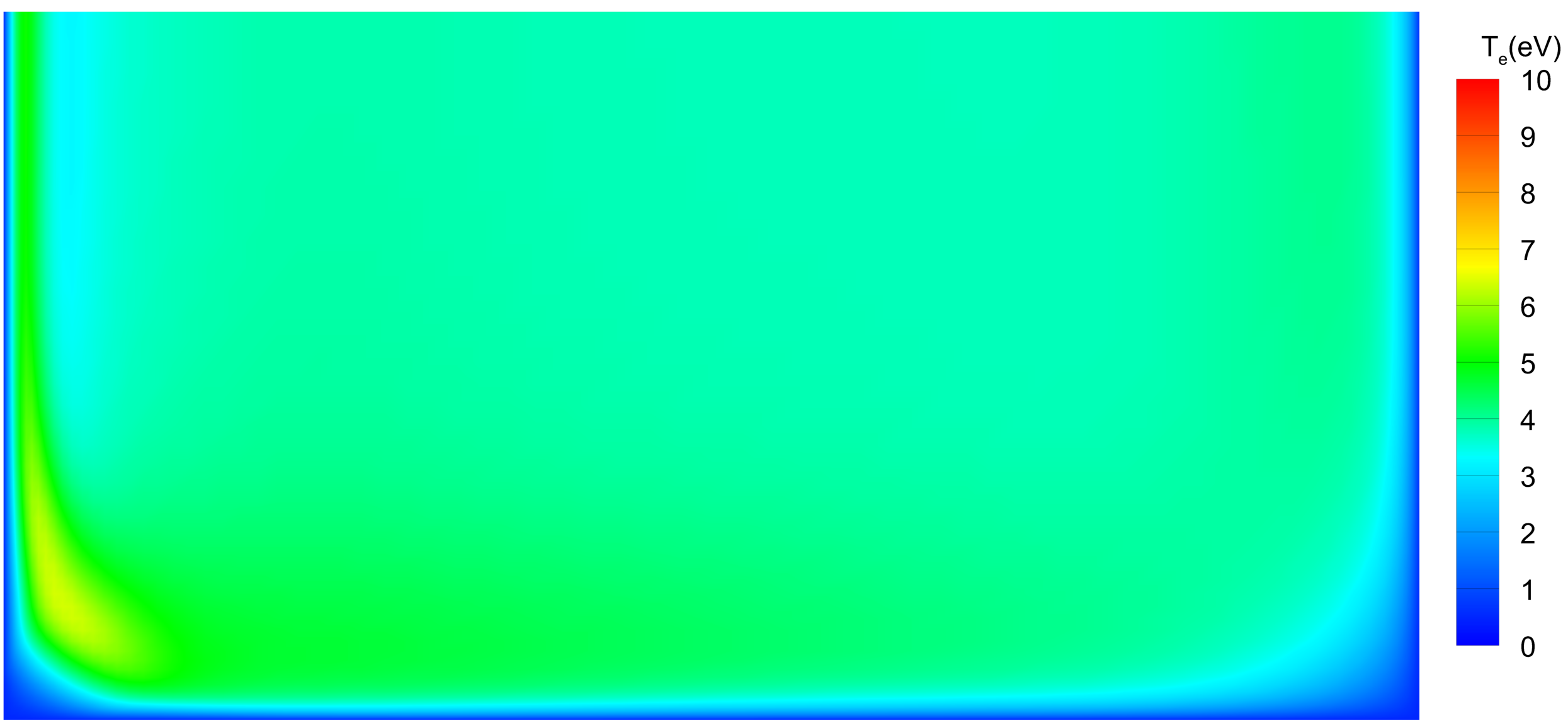}
    \caption{$t/T=0$}
\end{subfigure}
\hfill
\begin{subfigure}[b]{0.48\textwidth}
    \centering
    \includegraphics[width=\textwidth]{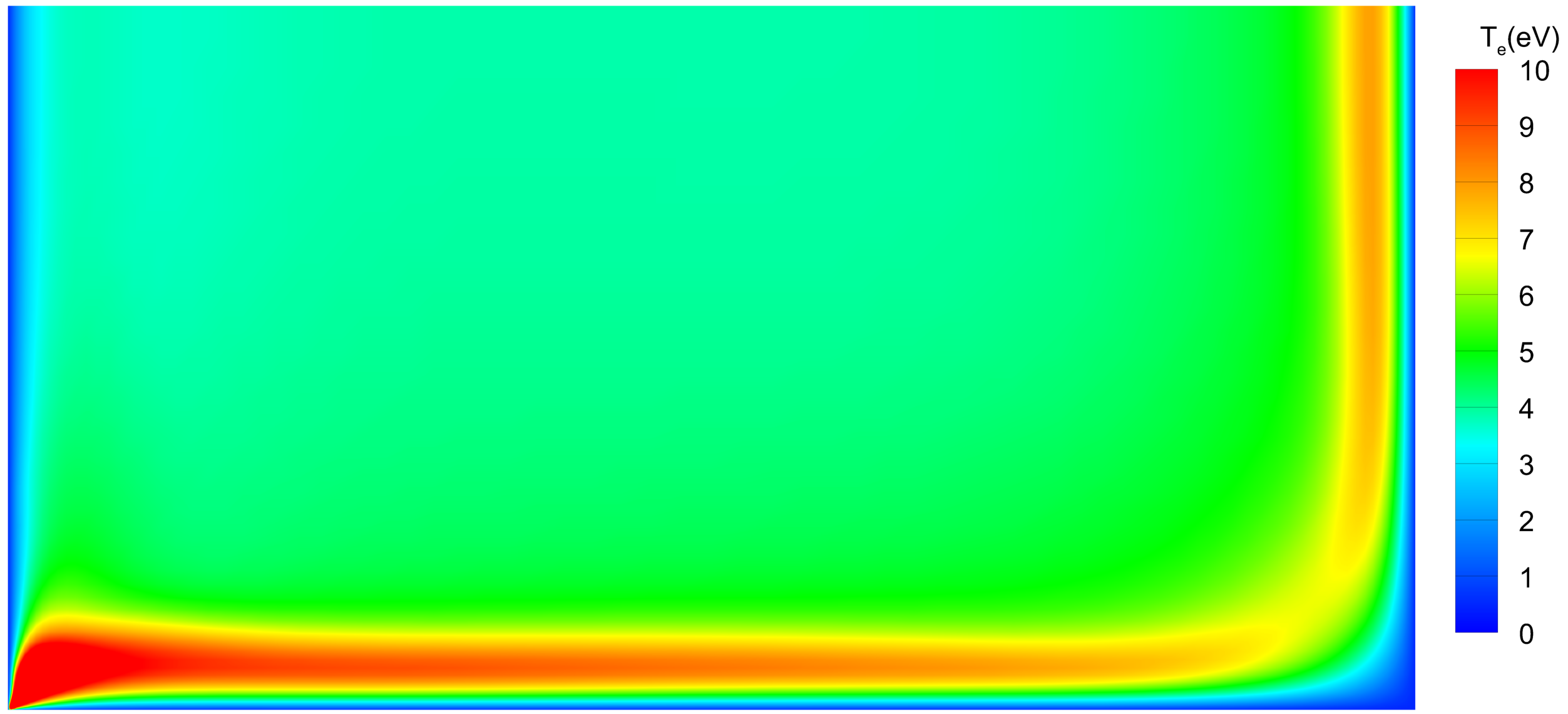}
    \caption{$t/T=0.25$}
\end{subfigure}

\vspace{1.0em}

\begin{subfigure}[b]{0.48\textwidth}
    \centering
    \includegraphics[width=\textwidth]{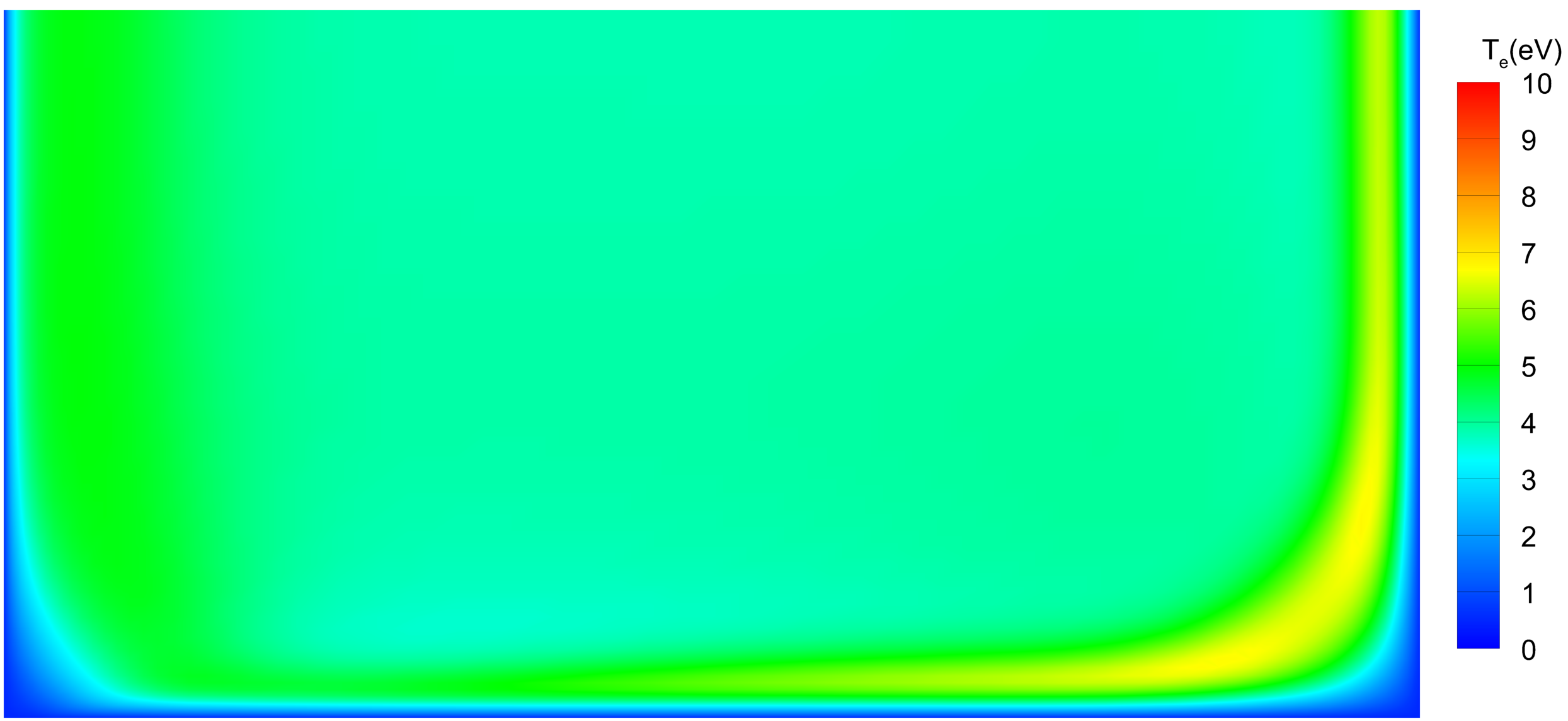}
    \caption{$t/T=0.50$}
\end{subfigure}
\hfill
\begin{subfigure}[b]{0.48\textwidth}
    \centering
    \includegraphics[width=\textwidth]{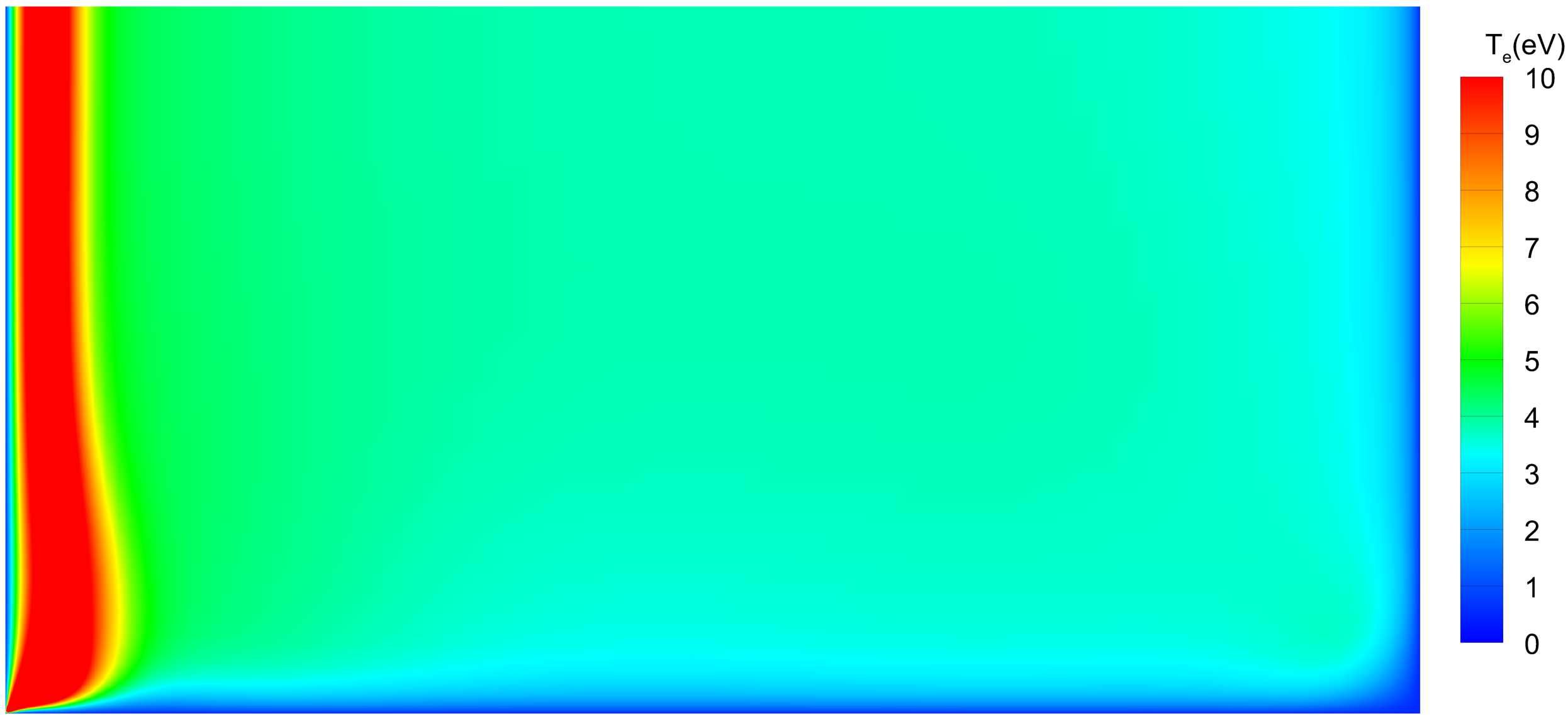}
    \caption{$t/T=0.75$}
\end{subfigure}
\caption{Phase-resolved contours of the electron temperature $T_e$ in Case 2.}
\label{fig:case2_Te_contours}
\end{figure}

Beyond the macroscopic density drop, the inclusion of the grounded wall significantly distorts the local discharge dynamics, leading to pronounced multi-dimensional effects. Figure~\ref{fig:case2_phi_contours} and Figure~\ref{fig:case2_Te_contours} present the phase-resolved contours of the electrostatic potential and electron temperature, respectively. 

Unlike the strictly parallel contour lines in Case 1, the electrostatic potential in Case 2 exhibits a severe multi-dimensional bending, particularly in the bottom-left region. This bending is a corner effect arising from the geometric intersection of the powered RF electrode ($x=0$) and the grounded transverse wall ($y=0$). The interaction between the longitudinal RF sheath and the transverse grounded sheath creates a complex, localized two-dimensional electric field. 

Consequently, the electron heating mechanisms are also spatially modulated. As shown in the electron temperature contours (Fig.~\ref{fig:case2_Te_contours}), the regions of intense electron heating---which typically coincide with the expanding sheath edges---are heavily distorted near the bottom corner. During the sheath expansion phases (e.g., $t/T=0.25$ and $0.75$), the high-temperature zones wrap around the corner, demonstrating that the electrons are simultaneously subjected to acceleration from both the longitudinal and transverse sheath electric fields. These visual and quantitative results confirm that the present dual-time solver is highly capable of capturing complex, genuinely two-dimensional plasma-sheath interactions.

\section{Conclusion}

In this work, a robust, low-storage implicit dual-time finite-volume framework has been developed for the efficient fluid simulation of RF CCPs. To overcome the severe numerical stiffness and prohibitive time-step constraints inherent in low-temperature plasma modeling, the proposed framework employs a BDF to strictly decouple the physical time advancement from explicit stability limits, while solving the resulting nonlinear system via pseudo-time iterations. A localized block-implicit relaxation scheme is implemented to handle the stiff transport and chemical source terms at the cell level, effectively circumventing the massive memory overhead typical of conventional fully implicit solvers. This implicit transport coupling is further combined with a semi-implicit treatment of Poisson's equation to accelerate the electrostatic coupling.

The numerical framework was systematically verified against a standard one-dimensional argon discharge benchmark. The verification studies confirmed that the implicit dual-time formulation allows the physical time step to be selected based solely on the required temporal accuracy. Through the optimal selection of the physical time step, pseudo-CFL number, and inner iteration limit, the macroscopic plasma dynamics were rapidly driven to a highly accurate periodic state within approximately $1000~\mathrm{s}$ of wall-clock time using a single-threaded execution, demonstrating the satisfactory computational efficiency of the algorithm.

To further demonstrate the multidimensional applicability of the proposed method, the framework was extended to genuine two-dimensional configurations incorporating a transverse grounded wall. The numerical results successfully captured the substantial depletion of the global plasma bulk induced by continuous transverse particle losses. Detailed phase-resolved contours revealed pronounced corner effects, accurately capturing the multi-dimensional distortion of the electrostatic potential and localized electron heating zones induced by the transverse boundaries. These findings prove that the proposed framework is highly capable of resolving complex, genuinely two-dimensional plasma-sheath interactions, thereby establishing a highly efficient and memory-friendly pathway for the predictive modeling of multi-dimensional low-temperature plasmas.

Building upon this robust time-domain framework, future research efforts will focus on broadening the solver's applicability to more complex discharge phenomena and larger computational scales. Ongoing developments aim to generalize the model for dual-frequency and multi-frequency CCP discharges, enabling the detailed study of the nonlinear coupling between low-frequency ion acceleration and high-frequency electron heating. As these physical investigations progressively scale toward full reactor-level simulations, accommodating the associated computational demands will become a primary focus. To this end, future work will integrate parallel computing architectures with advanced algorithmic accelerations, particularly exploring the adaptation of frequency-domain approaches, such as the harmonic balance method, to provide highly efficient convergence strategies for strongly driven periodic plasmas.

 \section*{Acknowledgments}
The current research is supported by National Science Foundation of China (92371107) and Hong Kong research grant council (16208324).

\bibliographystyle{elsarticle-num} 
\bibliography{elsarticle-num.bib}

\end{document}